\documentclass[structabstract]{aa}  
\usepackage{graphicx}
\usepackage[varg]{txfonts}
\usepackage{pifont}
\usepackage{bigstrut}
\usepackage{lscape}
\usepackage{rotating}
\usepackage{subfigure}
\usepackage{color}
\usepackage{natbib}            
\bibpunct{(}{)}{;}{a}{}{,}  
\begin{document}

\title{Synthetic photometry for carbon-rich giants}

\subtitle{IV. An extensive grid of dynamic atmosphere and wind models \thanks{Appendices are only available in electronic form at
\newline 
\texttt{http://www.aanda.org}}\thanks{Table B.1,  photometry and spectra for all snapshots are only available at the CDS. They can
also be downloaded from \newline\texttt{http://www.astro.uu.se/AGBmodels}} }

   \author{K.~Eriksson\inst{1}
           W.~Nowotny\inst{2}   \and
           S.~H\"ofner\inst{1}   \and                  
           B.~Aringer\inst{2}   \and
           A.~Wachter\inst{3} 
           }
          
   \institute{Department of Physics and Astronomy, 
              Division of Astronomy and Space Physics,
              Uppsala University,
              Box 516, SE-75120 Uppsala, Sweden\\
              \email{kjell.eriksson@physics.uu.se}
           \and   
              University of Vienna, Department of Astrophysics, 
              T\"urkenschanzstra{\ss}e 17, A-1180 Wien, Austria
         \and
             Departamento de Astronomia de la Universidad de Guanajuato,
             Apartado Postal 144, C.P. 36000, Guanajuato, GTO,
              Mexico
             }

\date{Received; accepted}

\titlerunning{Synthetic photometry for carbon-rich giants IV.}
\authorrunning{K. Eriksson et al. }

 
\abstract
{The evolution and spectral properties of stars on the asymptotic 
giant branch (AGB) are significantly affected by mass loss through dusty stellar winds. 
Dynamic atmosphere and wind models are an essential tool for studying these evolved stars, 
both individually and as members of stellar populations, to understand their contribution to 
the integrated light and chemical evolution of galaxies. }
{This paper is part of a series with the purpose of testing state-of-the-art atmosphere and wind models 
of C-type AGB stars against observations, and making them available to the community for use
in various theoretical and observational studies. }
{We have computed low-resolution spectra and photometry (in the wavelength range 0.35 -- 25 $\mu$m) 
for a grid of 540 dynamic models with stellar parameters typical of solar-metallicity C-rich AGB stars and 
with a range of pulsation amplitudes. 
The models cover the dynamic atmosphere and dusty outflow (if present), assuming spherical symmetry, 
and taking opacities of gas-phase species and dust grains consistently into account. 
To characterize the time-dependent dynamic and photometric behaviour of the models in a concise way 
we defined a number of classes for models with and without winds. }
{Comparisons with observed data in general show a quite satisfactory agreement for example regarding 
mass-loss rates vs. \mbox{($J$\,--\,$K$)} colours or $K$ magnitudes vs. \mbox{($J$\,--\,$K$)} colours. 
Some exceptions from the good overall agreement, however, are found and attributed to 
the range of input parameters (e.g. relatively high carbon excesses) or intrinsic model assumptions
(e.g. small particle limit for grain opacities). }
%
{While current results indicate that some changes in model assumptions and parameter ranges should 
be made in the future to bring certain synthetic observables into better agreement with observations, 
it seems unlikely that these pending improvements will significantly affect the mass-loss rates of the models. }

   \keywords{
             stars: AGB and post-AGB --
             stars: atmospheres --
             stars: carbon --
             stars: variables: general --
             stars: mass-loss --
             circumstellar matter
            }

   \maketitle


\section{Introduction}
\label{s:intro}

High luminosities, low effective temperatures and highly dynamic atmospheres are defining properties 
of AGB stars that make them popular targets for observations ranging from photometric monitoring 
to high-resolution spectroscopy and, more recently, NIR and MIR interferometry. 
They represent a crucial stage in the life of low- and intermediate-mass
stars where mass loss 
is a decisive factor for
the evolution, and newly produced chemical elements, most 
notably carbon, are fed into the surrounding ISM in the form of gas and dust by stellar winds. 
The production of new elements and their transport to the surface are the result of complex interlinked 
processes involving thermal pulses, hot bottom burning, and the third dredge-up
\citep[see, e.g.][]{Herw05, Kara11, StLa11, Marig13}.
In addition to being interesting objects in their own right, AGB stars are prominent members of 
stellar populations and important tools for extragalactic studies. 
All of these facts are good reasons for developing a better physical understanding of these stars, 
and a crucial part of that effort is the construction of detailed, realistic models of their atmospheres 
and winds. 

This article is the  fourth in a series dedicated to an in-depth analysis of synthetic observables, 
in particular,  low-resolution spectra and photometry, resulting from state-of-the-art model atmospheres 
for carbon-rich giant stars. The first paper \citep[][Paper~I]{AGNML09}
presented a large grid of  classical hydrostatic model atmospheres for C-type AGB stars, 
demonstrating that such models are applicable to the interpretation of objects with 
low pulsation amplitudes and very low mass-loss rates. 
For effective temperatures above 2800\,K these models agree very well with observations. 
For cooler, more dynamic stars, however, the formation of dusty circumstellar envelopes severely 
affects the spectral energy distribution. 
The time-dependent effects of shock waves and dust formation on observable properties were 
discussed in detail in the second paper of this series \citep[][Paper~II]{NoAHL11},
based on one dynamic model  with an intermediate mass-loss rate. 
The third paper \citep[][Paper~III]{NoAHE13} extended the study of 
dynamic models to a small sample with stellar parameters representing different stages of 
AGB evolution, resulting in a sequence of increasing mass loss and
correspondingly stronger circumstellar reprocessing of the photospheric fluxes. 
A detailed comparison with photometric observations of C-type AGB stars showed that our 
dynamic atmosphere and wind models reproduce a wide range of observed properties.  
References to other work based on hydrostatic and dynamic models can be found in 
Papers~I and II, respectively.

In this paper we present results for 540 dynamic models, to a large extent based on the grid by 
\cite{MatWH10}, to provide a consistent set of wind properties 
(mass-loss rates, outflow velocities, dust-to-gas ratios) and synthetic observables 
(low-resolution spectra, photometry) to the community. 
We discuss the influence of fundamental stellar parameters, pulsation properties, and certain 
model assumptions on the results, and compare them with various observations of carbon stars. 
This unique data set and accompanying information is made available in electronic form, 
providing input for theoretical studies on the evolution of individual stars and stellar populations, 
as well as for interpretations of observations.

Section \ref{s:modelling} gives a short summary of modelling methods, described in more detail 
in earlier papers, and an overview of the physical parameters defining the model grid. 
In Sect.~\ref{s:repcases} we define classes characterizing the time-dependent dynamic and 
photometric behaviour of the individual models in a compact form (with representative examples
discussed in Appendix~\ref{a:repremods}).
In Sect.~\ref{s:properties} we give an overview of the dynamic and photometric properties of 
the whole model grid, as well as a comparison with observations. 
Section~\ref{s:summary} presents a brief summary and our main conclusions.
Appendices~\ref{a:overviewdata} and \ref{a:download} contain a short description of the material
available at the CDS.

\section{Dynamic models and radiative transfer}
\label{s:modelling}

The modelling of dynamic structures and synthetic observables follows the 
approach described in Papers II and III of this series. 
Here we only give a brief summary, focusing on a few points of special interest, and referring 
to these and other papers for a more comprehensive description.

\subsection{Dynamic atmosphere and wind models}
\label{s:dynmodelling}

The time-dependent structures of atmospheres and winds are produced by  simultaneously solving 
the equations of hydrodynamics, frequency-dependent radiative transfer, and dust formation 
in spherical symmetry \citep[cf.][]{HoGAJ03, MatWH10}. 
The models cover two main dynamic aspects that affect the observable properties of AGB stars: 
pulsation-induced shock waves, and dust-driven winds (discussed in detail in Papers II and III). 
Each dynamic model is represented by a sequence of snapshots of radial structures, 
covering typically hundreds of pulsation periods.

The computations start from a hydrostatic, dust-free configuration characterized by the 
fundamental stellar parameters (mass, luminosity, effective temperature) and abundances of 
chemical elements, similar to classical model atmospheres. 
The effects of stellar pulsation are simulated by time-dependent boundary conditions at the inner edge 
of the com\-pu\-ta\-tion\-al domain, located just below the stellar photosphere. 
The velocity of the innermost mass shell is prescribed as a sinusoidal variation with period and 
amplitude as parameters (so-called piston model), and the amplitude of the accompanying 
luminosity variation can be adjusted separately (see below).

Stellar pulsation -- directly and indirectly -- causes deviations from equilibrium at various scales. 
On the macroscopic level, propagating shock waves are triggered in the atmosphere, leading to extended, 
variable atmospheric structures that do not resemble hydrostatic atmospheres at any phase \citep[Paper\,II,][]{NowHA10}. 
On microscopic scales, dust formation is proceeding far from equilibrium since the time scales for grain 
growth (involving the accumulation of 10$^6$ -- 10$^9$ atoms into a single grain) are 
similar to dynamic time-scales \citep{GHAGBbok}. 
Therefore both phenomena, that is, gas dynamics (including shocks) and non-equilibrium grain growth, 
are described by detailed time-dependent equations in our models. 
Regarding even smaller scales, such as molecular chemistry, models presented for example by Cherchneff and collaborators
\citep[e.g.][]{Cher06, Cher12}  indicate that the effects of propagating shocks may drive the abundances of molecules away from their 
chemical equilibrium values. 
Including of full non-equilibrium chemistry in dynamic atmosphere and wind models, however, would 
considerably increase the computational effort and is therefore currently beyond the scope of large model grids. 
The effects of assuming molecular equilibrium in the gas phase when computing grain growth rates (as done here 
and in other dynamic models in the literature) is expected to be moderate in C-type AGB stars, since the growth 
of carbon grains can in principle proceed via a range of C-bearing molecules (or C atoms) 
and does not critically depend on a particular species.

\subsection{Grid parameters}
\label{s:DMAs}

The stellar parameter combinations in the present grid  correspond to a sub-sample of  
the radiation-hydrodynamic model grid presented by \citet[][hereafter referred to as MWH10]{MatWH10},
with the constraint that they must be near the AGB  tracks for solar-abundance evolution models, 
like those of \cite{BeGMN08} or \cite{MGBGS08}.

The models considered here have fundamental parameters
(effective temperature, luminosity, and stellar mass) as given in Table~\ref{t:gridparam}. 
All models have solar abundances following \cite{AsGS05}, except for the carbon abundance.
The values are given on the scale where log N$_H$ $\equiv$  12.00. 
The abundances from \cite{AsGS05} correspond to a composition by mass, 
X/Y/Z = 0.73/0.25/0.015--0.020; the variation being due to the varying carbon abundance.
For every combination of stellar parameters in the table, models were computed with 
carbon excesses of log(C--O) + 12 = 8.2, 8.5 and 8.8                     
\footnote{
The relation between the carbon excess, log(C--O)+12, and the commonly
used quantity C/O is given in Table~\ref{t:coratios}. 
In contrast to other papers in the literature, we use
the carbon excess to characterize the models because this quantity directly translates into the amount
of carbon available for formation of carbon-bearing molecules (other than CO) and dust grains.
}
and with piston velocity amplitudes of $\Delta u_{\rm p}$ = 2, 4 and 6 km\,s$^{-1}$.
The pulsation periods are given by the luminosity through a period-luminosity relation following
\cite{FeGWC89}; for easy reference the periods are also given in Table~\ref{t:gridparam}
although they are not treated as an independent parameter. 
\footnote{
The $P-L$ relation by \cite{FeGWC89} is based on Miras in the LMC. 
In a diagram of $K$ magnitudes vs. period as, e.g., in \cite{Ita04}, our models would largely overlap
with the observed stars in sequence C. The stars belonging to this sequence are commonly
believed to be fundamental-mode pulsators \citep{Wood99}.
}

\begin{table}
\begin{center}
\caption{Combinations of fundamental stellar parameters covered by the model grid. 
Note that the period is not an independent parameter but coupled to the stellar luminosity, see text.
For each set of parameters listed here we varied the velocity and luminosity amplitude 
\bf
 ($\Delta u_{\rm p}$, $f_{\rm L}$) 
\rm
of the inner boundary.}
\begin{tabular}{cccll}
\hline
\hline
$T_\star$   & log $L_\star$  & $P$ & $M_\star$ & log(C--O)+12  \bigstrut[t] \\
 {[K]} &  [$L_\odot$] & [d] & [$M_\odot$]  &  [dex] \bigstrut[b] \\
\hline
2600 & 3.70  & 294 &  0.75, 1.0    & 8.2, 8.5, 8.8  \bigstrut[t] \\
&         3.85  &  390 & 0.75, 1.0, 1.5, 2.0  & 8.2, 8.5, 8.8  \\
&         4.00  &  525 & 1.0, 1.5, 2.0    & 8.2, 8.5, 8.8  \bigstrut[b] \\
\hline
2800 & 3.55  & 221 & 0.75    & 8.2, 8.5, 8.8  \bigstrut[t] \\
&         3.70  &  294 & 0.75, 1.0    & 8.2, 8.5, 8.8  \\
&         3.85  &  390 & 0.75, 1.0, 1.5, 2.0   & 8.2, 8.5, 8.8  \\
&         4.00  &  525 & 1.0, 1.5, 2.0   & 8.2, 8.5, 8.8  \bigstrut[b] \\
\hline
3000 & 3.55  &  221 & 0.75    & 8.2, 8.5, 8.8  \bigstrut[t] \\
&         3.70  &  294 & 0.75, 1.0   & 8.2, 8.5, 8.8  \\
&         3.85  &  390 & 0.75, 1.0, 1.5, 2.0   & 8.2, 8.5, 8.8 \\
&         4.00  &  525 & 1.5    & 8.2, 8.5, 8.8   \bigstrut[b] \\
\hline
3200 & 3.55  & 221 &  0.75   & 8.2, 8.5, 8.8  \bigstrut[t] \\
&         3.70  & 294 &  0.75, 1.0    & 8.2, 8.5, 8.8  \bigstrut[b]  \\
\hline
\end{tabular}
\label{t:gridparam}
\end{center}
\end{table}

During a pulsation cycle the radius of the innermost mass shell of the model and the luminosity at that position 
vary simultaneously, simulating a coupling between the variable brightness and size of the star. 
Without a quantitative model of the pulsating stellar interior, assumptions have to be made about the forms, 
the amplitudes, and the relative phases of these temporal variations. 
Regarding the dynamic boundary (i.e. gas velocity), the functional form is of minor importance since the 
outward-travelling waves quickly develop into shocks, and the kinetic energy transferred into the atmosphere 
is mostly determined by the velocity amplitude. 
Therefore we adopted the common assumption of a sinusoidal variation, parameterized by the velocity amplitude 
$\Delta u_{\rm p}$ and the pulsation period $P$. 
For simplicity, we assumed that the luminosity at the position of the innermost mass shell varies in phase with 
its radius, and that the amplitude of the luminosity variation is correlated with the radius amplitude. 

To avoid additional free parameters, we assumed in our earlier models \citep[e.g.][MWH10]{HoGAJ03}
 that the radius of the innermost model layer adjusts to the variable luminosity in such a way that the radiative flux 
 is constant in time. 
 This corresponds to the luminosity being proportional to the square of the radial position of the inner boundary. 
 However, it turned out that this tends to give too small photometric variations compared with observations. 
 Therefore we later generalized the description of the luminosity boundary condition to include an adjustable factor 
 $f_{\rm L}$, such that $f_{\rm L}$~=~1 reproduces the earlier results, and that the relative change in luminosity amplitude 
 compared with that case is directly proportional to $f_{\rm L}$  \citep[see,][for details]{NowHA10}.
In this study, we have varied the value of this parameter, using both the original value of $f_{\rm L}$~=~1 as in MWH10, 
and a newly calculated set of models with $f_{\rm L}$~=~2 (i.e. twice the luminosity amplitude of the original grid), 
these values are thought to bracket observed values (cf. Fig.~\ref{fig7}). 
In addition, several of the original models from MWH10 were recalculated over a longer time interval to ascertain 
their long-term behaviour. 
In total, 540 different dynamic model atmospheres are used here (229 of them produced winds and 311 did not 
lead to outflows).

\begin{table}
\begin{center}
\caption{Relation between the carbon excess measure log(C--O)+12 and the C/O ratio for the
adopted chemical composition \protect\citep{AsGS05}, i.e. using an oxygen abundance of 
log(N$_O$/N$_H$)+12 = 8.66.}
\begin{tabular}{cc}
\hline\hline
log(C--O)+12 & C/O \bigstrut[t]\bigstrut[b] \\
\hline
8.2 & 1.35 \bigstrut[t]\\
8.5 & 1.69 \\
8.8 & 2.38 \bigstrut[b] \\
\hline
\end{tabular}
\label{t:coratios}
\end{center}
\end{table}

\subsection{Dynamic properties}
\label{s:DMAdyn}

The resulting dynamic properties of these models are compiled in Table~B.1 
The listed values are time-averages of the mass-loss rates, outflow velocities,
and carbon condensation degrees, and were determined in the following way:
for a given model, the existence of a wind was defined as the outer boundary having reached 
25 stellar radii and the velocity of the gas at the outer boundary being positive (i.e. an outflow).
Then, for each (pulsation) cycle with the wind condition fulfilled, the mass-loss rate etc. were 
determined and later means and standard deviations were calculated. 
For most models, several hundred cycles were used to determine the means (at least a 
hundred cycles for the non-episodic models).

In general, mass loss is favoured by low $T_\star$ (easier to form dust at cooler temperatures),
high luminosity (more momentum transferred from radiation to dust grains), small stellar mass
(shallower potential well), large (C--O) (more free carbon to form amorphous carbon grains), 
and a large piston velocity amplitude (stellar layers reach out to a greater distance from the 
centre of the star).

\subsection{Synthetic spectra and photometry}
\label{s:synthspecphot}

For each combination of input parameters, snapshots of the atmospheric structures at various 
phases (typically 20 per period), during several cycles (at least four) were
used for the detailed {\it a~posteriori} radiative transfer calculations. 
We applied our opacity generation code {\tt COMA}
\citep[][and Paper~I]{Aring00} to compute atomic, molecular, and amorphous carbon (amC) dust 
opacities for all atmospheric layers and every wavelength point, assuming LTE. 
The dust opacity was calculated under the assumption of the small particle limit.
Note that the treatment of the opacities is consistent with that in the generation
of the dynamic models, except for a higher spectral resolution in the {\it a posteriori}
radiative transfer calculations (10\,000).
 
The resulting synthetic spectra in the wavelength range 0.35\,--\,25~$\mu$m were used to 
compute filter magnitudes in the Johnson-Cousins $BVRI$ system \citep{Besse90} 
and the Johnson-Glass $JHKLL'M$ system \citep{BB88}. 
More details can be found in previous articles, e.g. Paper~II.
The results given in Table~B.1    
include temporal mean of the $V$ and $K$ magnitudes, as well as
their (maximum) ranges, and also the mean values of the colours 
\mbox{($V$\,--\,$I$)}, \mbox{($V$\,--\,$K$)}, \mbox{($J$\,--\,$H$)} and \mbox{($H$\,--\,$K$)}.
Data for the other computed filters, as well as spectra, for all computed phases are available 
in electronic form, see Appendix~\ref{a:download}.

\begin{figure*}
  \includegraphics[width=17cm]{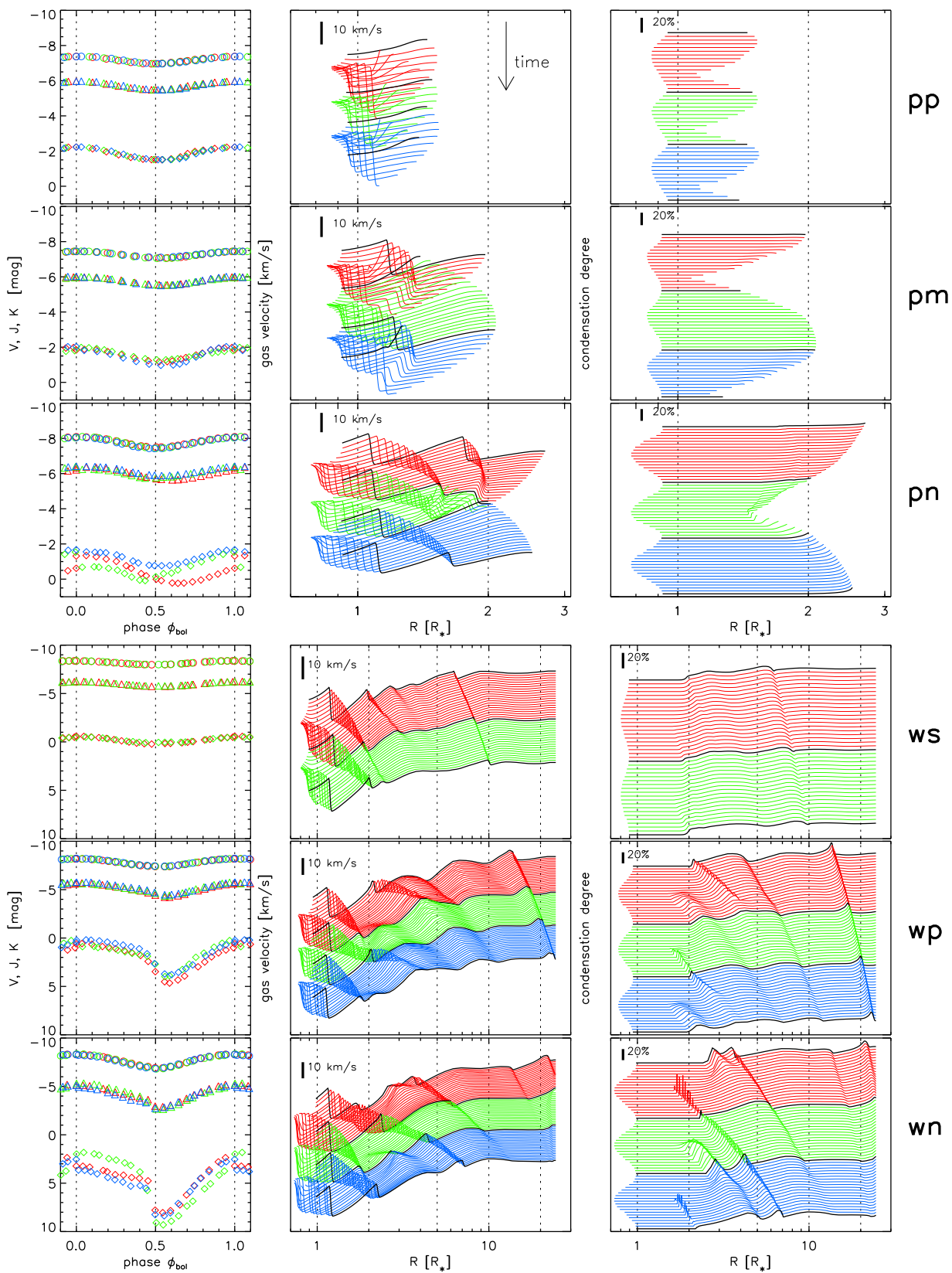}      
     \caption{Photometry and radial structures for representative models (see Table\,\ref{t:phototypic}) 
     without wind 
     (upper 3 rows, classes {\it pp}, {\it pm}, {\it pn}) and for models with winds (lower 3 rows,
     classes {\it ws},{\it wp}, {\it wn}). 
     The leftmost panels show the 
     brightness in the $V$ (diamonds), $J$ (triangles) and $K$ (circles) filters
     as a function of bolometric phase (with phase $\phi_{\rm bol}$=\textit{0.0} at maximum  light). 
     Points belonging to the same pulsation cycle have
     the same colour. Note that the photometric scales are different for models with 
     and without winds. Middle panels show radial profiles of the gas velocity. 
     Each curve represents a snapshot in time, the first at the top and later ones displaced by a 
     fixed amount downwards. The colours denote the different pulsation cycles and
     correspond to the colours in the leftmost panels. Black curves 
correspond to
 $\phi_{\rm bol}$=\textit{0.0}.
     Vertical dotted lines mark the distances 1, 2, 5, 10, and 20 stellar radii from the stellar centre. 
     A velocity range of 10 km/s is marked with bars. 
     The rightmost panels show radial profiles of the condensation degree of carbon plotted in the same 
     way as the velocities, here the bars mark a condensation degree of 20\%. }   
     \label{fig1rev}  
\end{figure*}

\section{Classes of dynamic and photometric behaviour}
\label{s:repcases}

The dynamic response of the atmosphere to periodic driving by the piston and to outward 
acceleration by radiative pressure on dust can lead to a complex time-dependent behaviour 
of the models (including their photometric properties).
This is because pulsation, atmospheric dynamics, and grain growth each have 
their own times cales. 
In Papers II and III of this series, we dealt with individual models where we could discuss the 
time-dependence of structures and observables in some detail. 
Here, we are faced with the challenge to characterize 540 models in a concise form, 
often with the use of temporal means of relevant quantities.

\subsection{Definition of classes and examples}

In this section, we define classes as a simple shorthand notation for describing the 
time-dependent (dynamic and photometric) behaviour of the models.
The models without a wind are divided into the classes {\it pp}, {\it pm}, and {\it pn}
(with the first letter {\it p} denoting ``pulsating atmosphere''). 
The  {\it pp} models are regularly {\it periodically pulsating}, 
repeating their dynamic behaviour every period.
In the {\it pm} class we put models that repeat after two (sometimes more) 
periods ({\it pulsating, multi-periodic}).
Other, more irregular models, are put in the class {\it pn} (pulsating, non-periodic).
Models with winds are similarly divided into the classes {\it ws}, {\it wp}, and {\it wn}
(with the first letter {\it w} denoting ``wind''). 
Here the {\it ws} class includes models with a {\it steady} wind with small temporal variations in 
mass-loss rates and wind velocities. 
The models in the {\it wp} class show {\it periodic} variations in the wind properties, while the {\it wn}
class includes models with more irregular behaviour ({\it non-periodic}). 
Finally, a transition  class {\it we} contains the {\it episodic} models, that is, models that show 
an intermittent outflow.

The time-dependent behaviour of the different classes (except {\it we})
is illustrated in Fig.~\ref{fig1rev} for representative models. 
The  middle and right panels in  the figure show radial profiles of  the gas velocity and 
the condensation-degree for carbon, respectively.  
Each curve represents a snapshot in time, with time running downwards and with 
different colours to denote different stellar pulsation cycles. 
The panels in the left column display light curves as a function of 
phase (with phase $\phi_{\rm bol}$=\textit{0.0} at maximum  light) in the photometric 
$V$, $J$ and $K$ filters (with the colour of the symbols denoting the different cycles).
Information on which specific models were chosen as representative cases, and more details about 
them can be found in Appendix~\ref{a:repremods}.    

It should be pointed out that dust formation is a necessary but not a sufficient condition
for the development of a wind, and
even for the models without a wind the presence of dust can be important (see Fig.~\ref{fig1rev}, 
the {\it pn} case).
Photometrically, the amC dust opacity, being roughly inversely proportional to the wavelength, 
will extinct the visual light much more than, for example, the near-IR radiation, and for light with
wavelengths longer than $\sim$\,2\,$\mu$m the dust emission will dominate absorption.

For the models without a wind in Fig.~\ref{fig1rev} (upper 3 panels) we see that even very small 
amounts of dust (corresponding to condensation degrees of the order of a few per mille
of carbon atoms) will noticeably affect the $V$ light-curve (e.g. the {\it pm} model), and that the 
position in time of minimum light depends on where and when the dust formation takes place 
(e.g. the {\it pn} model), especially for the $V$ filter.
The photometric behaviour of wind models (Fig.~\ref{fig1rev}, lower 3 panels) 
is also clearly related to dust formation. 
The outward-moving shocks with enhanced dust are quite noticeable. 
Moreover, the $V$ magnitude behaviour is clearly related to the amount of dust formed at 1.5 -- 2 
stellar radii around the luminosity minima.
Many models in class {\it ws} produce much dust, with a rather large but almost constant 
condensation degree of carbon.
These models then show comparatively small variations in the $V$ magnitude.
Finally, we find examples of models with photometric behaviour in $V$ that suggest a different 
period than that of the piston. 
An example is given in Fig.~\ref{factsheetfig}, where we see that one could derive
a period from the $V$ magnitudes that is twice the luminosity period (or the one derived 
from $K$ magnitudes). 
Another example of this behaviour, illustrated with detailed plots, can be found in Paper~III, 
Appendix C.
So again we see the importance of dust formation for the photometric
behaviour in the visual wavelength region.
\footnote{
It is interesting to speculate if this behavior (i.e. a dust formation cycle corresponding to two pulsation periods) 
explains observed stars lining up along Sequence D in a $K$ magnitude vs. period dia\-gram \citep[e.g.][]{Ita04}, 
with $K$ magnitude unaffected but observed periods corresponding to twice the fundamental mode pulsation. 
However, we would like to defer a discussion of this point to a future paper since current synthetic $V$ light 
curves 
are affected by the assumption of small particle limit for grain opacities. 
}

One model shows unusual photometric properties for its class. It has the parameters
$T_\star$ = 2600\,K, log\,$L$ = 3.70, one solar mass, log(C--O)+12 = 8.2, $\Delta u_{\rm p}$ = 6, and
$f_{\rm L}$ = 2 and belongs to the {\it pn} class. The effective temperature is low but the
luminosity and/or carbon excess is not high enough to initiate even a marginal wind.
Instead, the model consists of a dusty outer atmosphere that extends to 5--10 stellar radii 
and with a high carbon condensation degree of 40--80\%. Therefore, it is found in 
positions unexpected for its class in colour-magnitude and two-colour diagrams 
(Figs.~\ref{fig6} and \ref{fig8}).

\subsection{Trends with model parameters}

\begin{figure}
  \resizebox{\hsize}{!}{\includegraphics{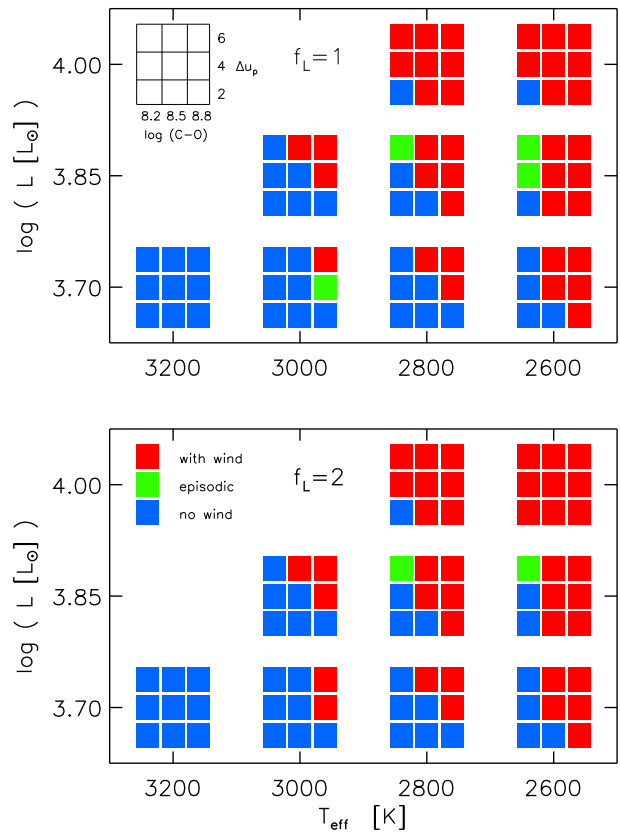}}                 
  \caption{Schematic overview showing the dynamic behaviour of the 1~$M_\odot$ models 
  as a function of other model parameters. Note that the temperature increases to the left in a 
  format resembling an HR-diagram.}
  \label{figifwind}
\end{figure}

The different kinds of dynamic behaviour can be understood in relation to the model parameters
as follows.

First, for a given combination of stellar mass, luminosity, and effective temperature, we note
that  with increasing values for the piston velocity amplitude 
($\Delta u_{\rm p}$ =  2 \ding{213} 4 \ding{213} 6 km s$^{-1}$) the 
outer stellar atmosphere will ballistically expand to increasingly greater heights. 
After reaching the maximum height, the downward moving gas will, at some height, 
encounter the outward moving gas in the next pulsation cycle. 
Depending on where this occurs, we see different scenarios: 
from cases where the whole atmosphere moves together in phase with the pulsation, to situations
where the outgoing shock wave interacts with down-falling material in complex ways.
So, we have cases from the {\it pp} class (where the behaviour repeats each period) over 
the {\it pm} class (with a repetition after two or more periods), to the {\it pn} class with 
a more irregular behaviour.
If the outward moving gas becomes cool and dense enough to allow dust formation, 
this will affect both the dynamic and photometric behaviour. 
Dynamically, by the additional momentum given to material through
the radiative acceleration of the dust and then to the gas via the collisional coupling. 
If the radiative acceleration exceeds the gravitational one, a stellar wind will form. 
We find that, when dust forms, this occurs around 1.5 -- 2 stellar radii (from the stellar 
centre) and outwards.
Depending on how much dust forms and where it forms, we find different kinds of wind classes:
for the {\it ws} class we see small variations of the degree of carbon condensation into dust at the same
distance, for models in the {\it wp} class dust formation occurs in a regular way
(similar amounts of dust at a similar distance every cycle or in a periodic pattern),
and for the {\it wn} class models the amounts and positions of dust formed vary in a more
irregular way. 

Second, for a given combination of stellar mass, luminosity and effective temperature, we find
that with increasing carbon excess (log(C--O)+12 = 8.2 \ding{213} 8.5 \ding{213} 8.8) 
the dust-to-gas ratio increases: a high carbon excess facilitates the formation of a wind (and, 
as we will see, also increases the terminal wind velocity through increased radiative acceleration), 
as well as increasing the effects of dust on the photometry. 

The behaviour as a function of stellar mass, luminosity and effective temperature can be
understood in the following way: (i) dust formation is favoured by a lower temperature, 
and  (ii) a higher luminosity or a lower stellar mass will help the radiative acceleration on dust
to overcome the stellar gravity.
A schematic representation of the occurrence of stellar winds for 1~$M_\odot$ models is given
in a format like an HR-diagram in Fig.~\ref{figifwind}, illustrating the trends discussed above.

\begin{figure}
  \resizebox{\hsize}{!}{\includegraphics{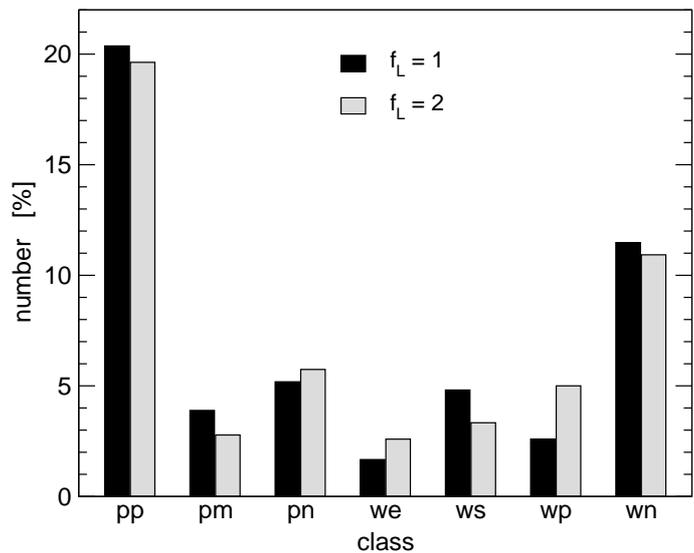}} 
  \caption{Distribution of models over classes of time-dependent behaviour, shown
  separately for the two values of $f_{\rm L}$.}
  \label{clsdistr}
\end{figure}

The parameter $f_{\rm L}$, discussed in Sect.~\ref{s:DMAs}, regulates the luminosity amplitude
of the model without changing the kinetic energy input due to pulsation (which is determined by
$\Delta u_{\rm p}$ and $P$).
Figure~\ref{clsdistr} where the distribution of our models over  the different classes is shown, separately for 
the different $f_{\rm L}$ values,  gives a first indication that the effect of $f_{\rm L}$ on the 
dynamics and wind properties is minor. 
The two cases give similar distributions, the only possibly significant difference between
$f_{\rm L}$ = 1 and $f_{\rm L}$ = 2 being that the relative numbers of models in the classes {\it wp} 
and {\it ws} change.
We further discuss the effects of $f_{\rm L}$ in a more quantitative way in Sect.~\ref{s:windeffects}.

\section{Overall properties of the model grid}         
\label{s:properties}

After looking into representative examples of time-dependent behaviour we now
discuss the model grid as a whole, that is, trends of resulting properties with input parameters
and how our group of models compares to observational samples.
As mentioned in Sect.~\ref{s:DMAs}, the present grid parameters form a sub-sample of 
the original grid range of MWH10.
They cover the region where most  carbon-star models appear 
in synthetic stellar evolution calculations for solar metal abundances (e.g. those 
of \cite{BeGMN08} or \cite{MGBGS08}).

Directly from the radiation hydrodynamic calculations we obtain the mean mass-loss rates, 
wind velocities, carbon condensation degrees, and dust-to-gas ratios as well as their
standard deviations (everything computed at the outer boundary at 25 stellar 
radii, see Sect.~\ref{s:DMAdyn}). 
The {\it a posteriori} radiative transfer calculations give us the mean photometric magnitudes and 
their variations in different filters (cf. Sect.~\ref{s:synthspecphot}).
Here we restrict the discussion to temporal means of all values (usually taken over several 
hundreds of periods).

When we compare the results from the present calculations with data from
various observational samples of carbon stars, it is important to bear in mind 
that all parameter combinations get the same weight in our grid, whereas in the observational samples
the number of targets with corresponding stellar parameters will depend on 
the initial mass function, stellar evolution effects, observational selection criteria, etc. 

In this section we first discuss  the dynamic properties of the grid and then present the
photometric results from the radiative transfer calculations.

\subsection{Wind properties}          
\label{s:windeffects}

\begin{figure}
  \resizebox{\hsize}{!}{\includegraphics{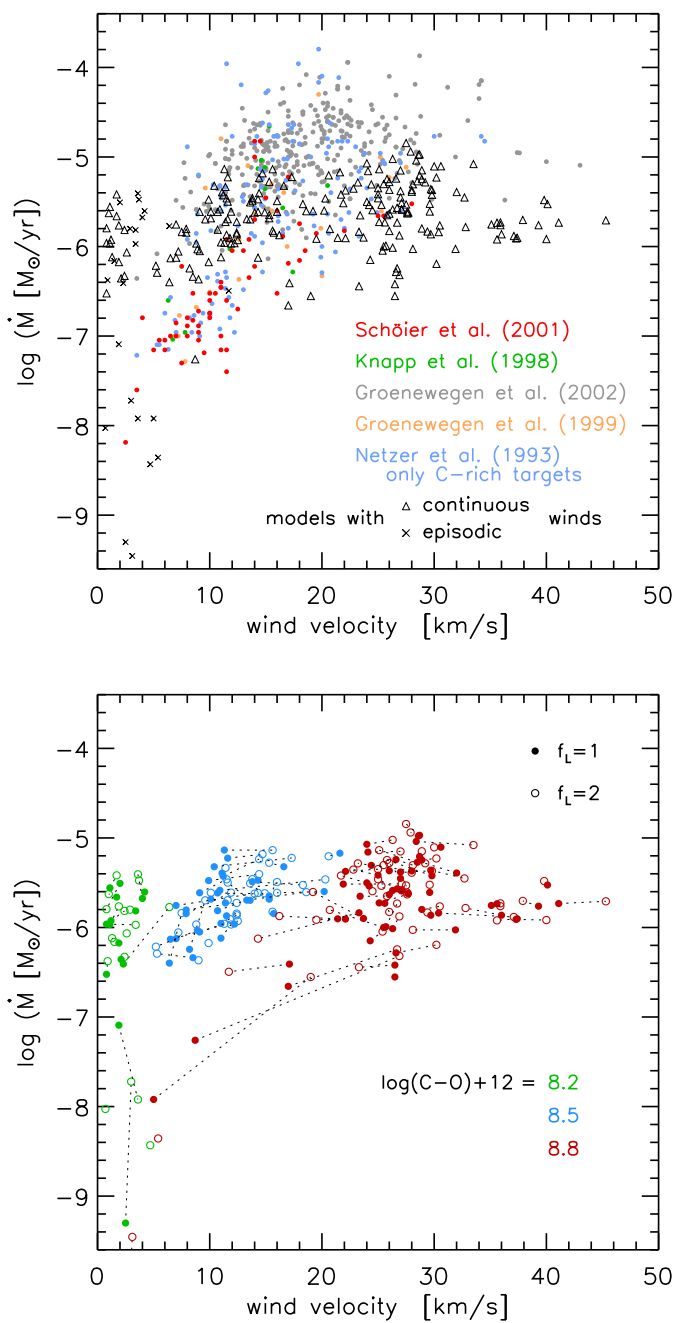}} 
  \caption{Mass-loss rate plotted versus the wind velocity for all models
   with outflows. In the upper panel the triangles mark models with
  steady winds while the crosses denote models with episodic winds. The other symbols with 
  different colours show  values based on observations
  of various C-rich giants,
  adopted from different sources as given by the legend.
  The  lower panel shows the values for the
  models which have winds for both $f_{\rm L}$~=~1 and $f_{\rm L}$~=~2. The latter are plotted as
  open circles and models with identical parameters except $f_{\rm L}$ are connected. The different
  colours denote various carbon excesses.}
  \label{fig2}
\end{figure}

Mass-loss rates and outflow velocities are two essential wind properties, and are also direct results of 
the dynamic models. 
In Fig.~\ref{fig2} (upper panel) we show a comparison of these quantities with empirical data 
adopted from the literature.
The dependence of model properties on carbon excess and luminosity amplitude is shown in the lower panel:
in the current grid there are 229 models, i.e. 42\% of the models, that give outflows.
In general, the wind models fall nicely within the region covered by observed stars.
However, there is one noticeable  exception: a group of models with very low wind speeds 
and relatively high mass-loss rates, corresponding to the lowest log(C--O)+12 value (8.2),
which have no counterparts among the observed stars. 
These models can probably be brought into better agreement with observations by giving up the
small-particle limit assumption when computing grain opacities (used for keeping the models
comparable to MWH10). 
An exploratory study by \cite{MatH11} demonstrated that the outflow speeds for models with relatively 
slow winds tend to increase considerably when using proper size-dependent 
grain opacities while the mass-loss rates are largely unaffected 
(see the second column of Fig.4 in their paper). 
This effect is due to a  higher efficiency factor for grains with radii of a few tenths of a micron 
\citep[see Fig.3 in][]{MatH11} and would most likely shift the group of low 
carbon excess models to the right in Fig.~\ref{fig2} (note that there seems to be little or 
no effect on models with wind velocities higher than 10 km\,s$^{-1}$ with small-particle limit opacities).

As already stated in MWH10, we note a lack of models with low mass-loss rates,
except for a few cases with episodic winds.
We also note in Fig.~\ref{fig2} that some models with log(C--O)+12~=~8.8 show higher wind velocities 
than the bulk of the observed stars,  probably indicating that such high carbon abundances are rare.
Furthermore, the current model grid does not cover objects with the highest mass-loss rates
derived for instance by \cite{GSSP02} or \cite{NE93}. 
This might be due to the selected physical parameters or uncertainties in the empirical mass-loss rates.

For the comparisons with empirical data it should be kept in mind that mass-loss rates 
are direct results of the models, whereas they may be less straightforward 
to derive from observations, depending on the type and quality of data, as well as the methods
of interpretation.
\footnote{
On the other hand, certain directly observable quantities may not put strong constraints on the dynamic 
models if they measure parameters that have little influence on the resulting model properties 
(see Sect.~\ref{s:coloureffects}). 
Often, comparisons are therefore a trade-off between direct accessibility and relevance of 
the information, and the choice of quantities to be compared between models and observations 
is non-trivial.
} 
\cite{RSOL08} concluded that the mass-loss rates in \cite{SchoO01} are reliable within a factor of three.
Values derived with other methods are probably less accurate.
For the empirical data shown in Fig.~\ref{fig2} we checked that for stars that appear in more than 
one sample, the derived mass-loss rates  do not deviate significantly from the values given by \cite{SchoO01}.

We now examine the role of the parameter $f_{\rm L}$,  which was introduced to bring the light curve
amplitudes into better agreement with observations without changing the input of kinetic energy due 
to pulsation (see Sect.~\ref{s:DMAs}).
In the grid by MWH10 this parameter was held constant at $f_{\rm L}$ = 1.
We find that the dynamic effects of changing this parameter are minor, while the effects on 
the photometric variations are quite pronounced.
In Fig.~\ref{fig2} we see that when $f_{\rm L}$ is increased from 1 to 2, the mass-loss rates are 
basically unchanged (except for marginal winds, i.e. low mass-loss rates) and that 
changes in the outflow velocity go both ways although a velocity increase is 
slightly more common ($\sim$\,60\%). 
Therefore, we conclude that it should be safe to use the mass-loss rates from
the MWH10 grid for example in stellar evolution calculations.

\begin{figure}
  \resizebox{\hsize}{!}{\includegraphics{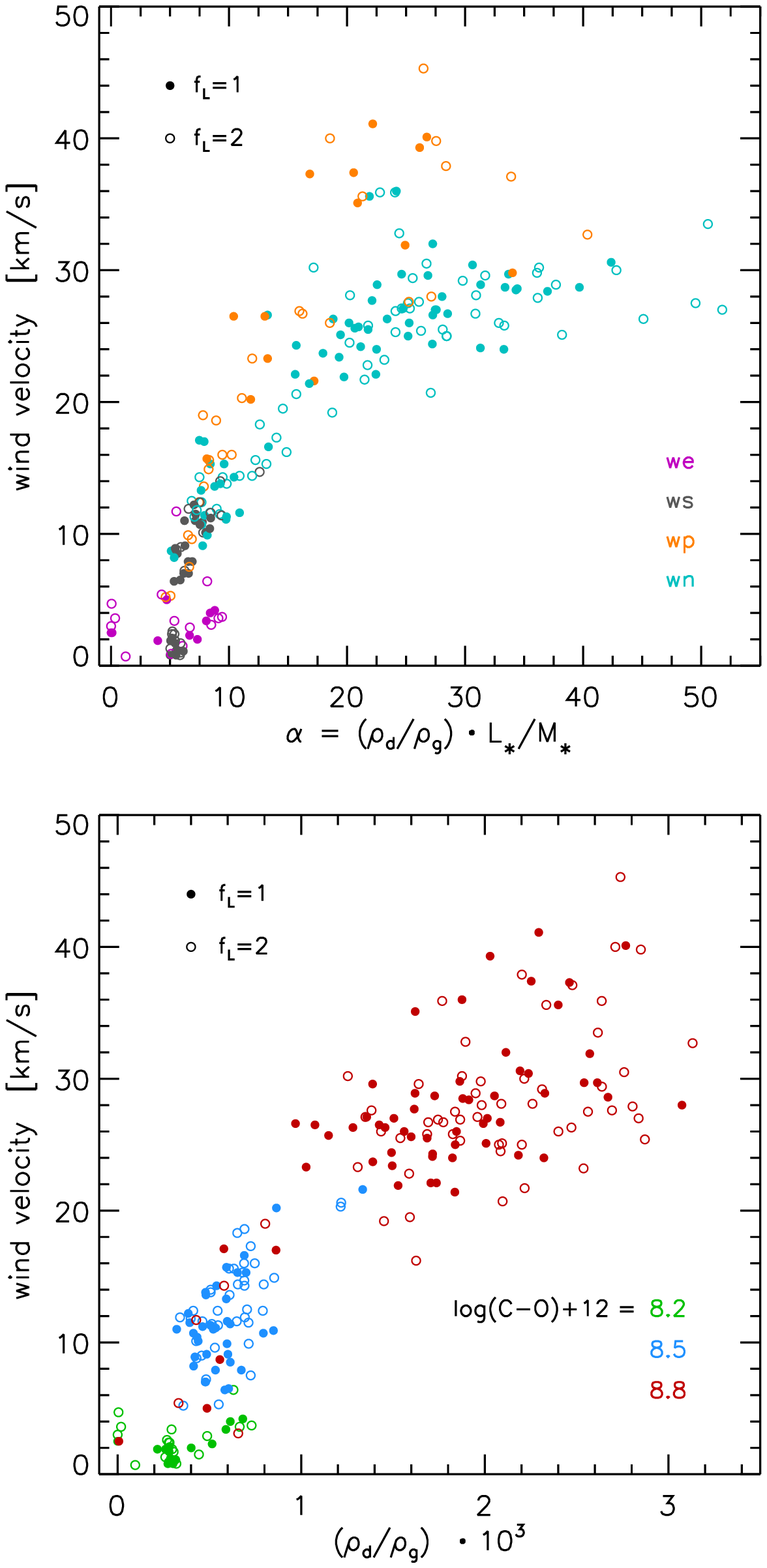}}  
  \caption{The terminal wind velocities as a function of $\alpha$ (upper panel) and 
  as a function of the dust-to-gas ratio (lower panel) for all models developing outflows.}
  \label{fig3}
\end{figure}

\cite{MatWH10} studied the dynamic properties as a function of 
a quantity $\alpha$, defined there as $(\rho_d/\rho_g)(L_\star/M_\star)$,
which is proportional to the ratio of radiative to gravitational acceleration.
As discussed there, one expects that the outflow velocities are more directly correlated 
with $\alpha$ than the mass-loss rates.
In the upper panel of Fig.~\ref{fig3} we plot the outflow velocities vs. $\alpha$ for our grid, where 
the different dynamic classes are marked.
Excepting episodic models, we see a relatively narrow relation up to about an outflow velocity 
of 20 km\,s$^{-1}$, and a tendency for a
separation for higher velocities such that the class {\it wp} continues 
the steeper trend and for other classes the increase with $\alpha$ is much slower, 
leveling off at a terminal velocity around 30 km s$^{-1}$. 
The reason could be a constructive resonance between dust formation and radiative 
acceleration for the {\it wp} class models (since the radiative acceleration of the dust varies 
with time as the luminosity).
Note that the corresponding figure in MWH10 looks different because of their
selection of stellar parameters (they also included models with log L = 4.15 and log(C--O)+12 = 9.1),
and because they did not have any models with $f_{\rm L}$~=~2.

To separate the effect of the dust-to-gas ratio on the outflow velocities
from the influence of stellar parameters, see also the lower panel of Fig.~\ref{fig3}.
The range in $\alpha$ is mainly a range in $(\rho_d/\rho_g)$ as $(L_\star/M_\star)$ only
varies by a factor of about three for the models. 
The figure shows that the outflow velocity dependence on the  $(\rho_d/\rho_g)$ ratio 
largely is an effect of the carbon excesses (as a large carbon excess favours the
condensation of amorphous carbon dust and consequently facilitates the 
acceleration of gas and dust).
The variation of the $(\rho_d/\rho_g)$ ratio by a factor of 30 or more found in the models makes 
it questionable if the commonly used assumption of a constant $(\rho_d/\rho_g)$ value 
when deriving mass-loss rates from observations is justified.

\begin{figure}
  \resizebox{\hsize}{!}{\includegraphics{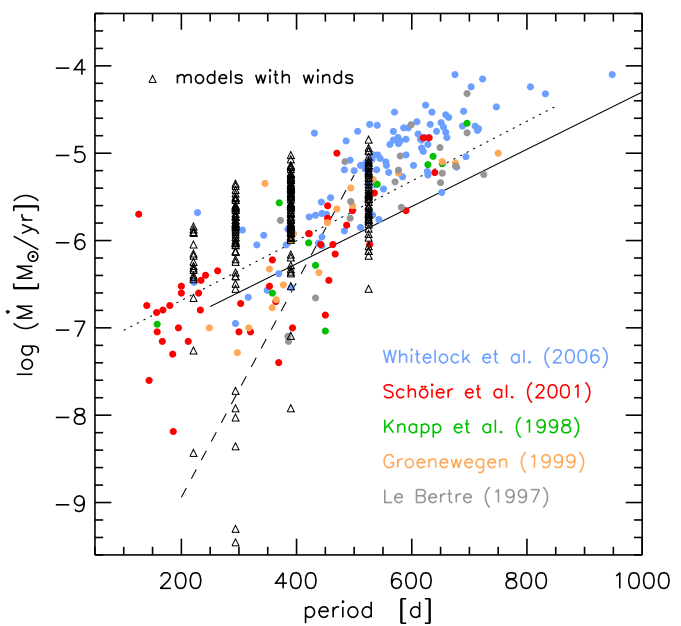}}   
  \caption{Mass-loss rate as a function of pulsation period for the models with winds,
     plotted as black triangles compared to observations of various galactic C-rich targets
     from various sources. 
     The plotted relations come from 
     \protect\cite{VW93} (dashed line), 
     \protect\cite{GroeWSK98} (their Eq. 9, C-miras; solid line), and  
     \protect\cite{dBDKJVKM10} (fit on their page 21; dotted line).}
  \label{fig3b}
\end{figure}

The mass-loss rates derived from several observational studies are plotted against observed periods 
in Fig.~\ref{fig3b} and are compared to the models. 
As mentioned above, the pulsation periods for our models are taken from an adopted 
period-luminosity relation. 
We see here that the slopes of the mass-loss rate vs. period relation from our computations are
fairly consistent with the empirical relations by \cite{GroeWSK98} and \cite{dBDKJVKM10}. 
The relation from \cite{VW93} seems too steep, again bearing in mind that 
our grid does not correspond to a stellar population. 
The computed mass-loss rates for the models with the shortest periods (i.e. lowest luminosities) 
are higher than those for most observed short-period AGB stars (these models usually have high 
carbon excess and large piston amplitudes and have no observed counterparts in this plot).

\begin{figure}
  \resizebox{\hsize}{!}{\includegraphics{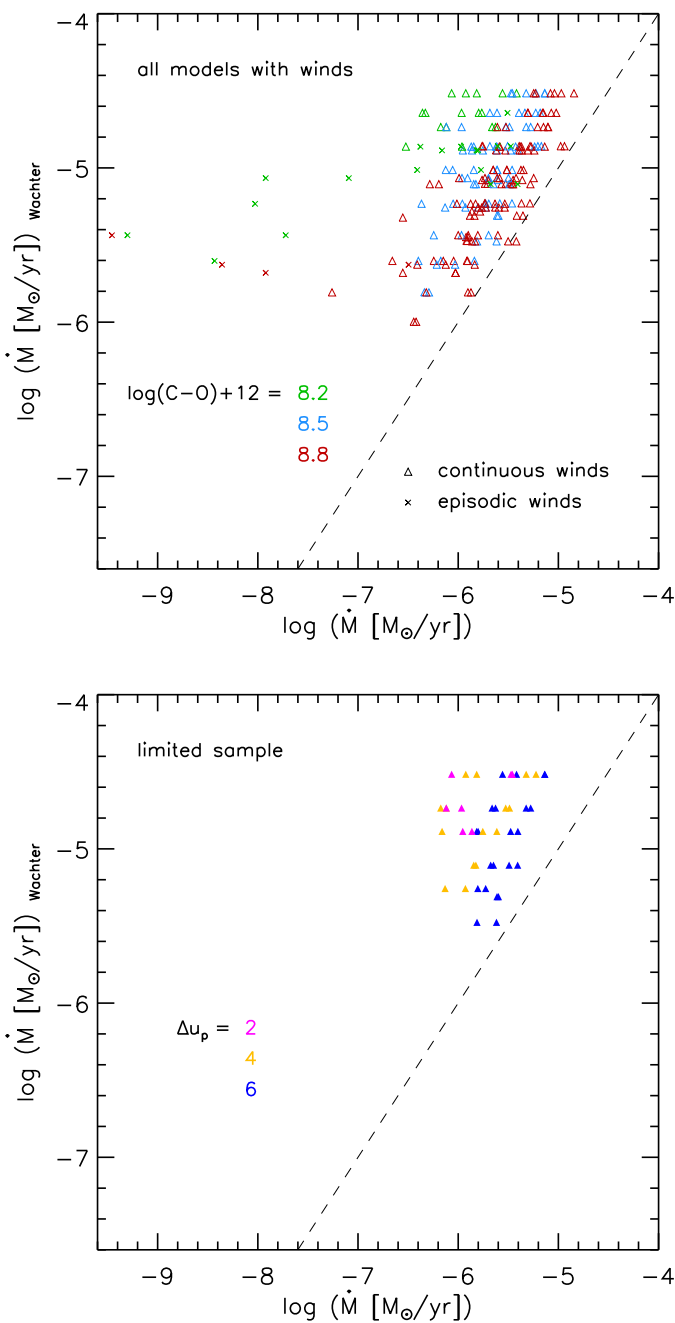}}   
  \caption{
  Comparison of the mass-loss rates for the present model grid
  with the values computed from the mass-loss rate formula of \protect\cite{Wacht02} using the
  same stellar parameters.
  The top panel shows the result for all models with winds, while the lower panel shows it for
  a subsample of models that fall within the range for which the formula was originally derived
  (log(C--O)+12 = 8.2 and 8.5 and one solar mass). 
  The systematic differences are due to differences in the underlying models, see text. 
  }
  \label{figWacht}
\end{figure}

Finally, in Fig.~\ref{figWacht}, we also compare our model grid with the mass loss formula given by 
\cite{Wacht02},
based on an older generation of dust-driven wind models.
The mass-loss rates resulting from our models are systematically, and often significantly, lower than 
the values obtained from the mass-loss formula for the same stellar parameters. 
The main physical difference between the two sets of models is the treatment of gas opacities and radiative transfer. 
In contrast to the dynamical models discussed here (which include frequency-dependent radiative transfer 
with detailed gas and dust opacities, see Sect.~\ref{s:dynmodelling}), the models used by \cite{Wacht02} are based on 
grey radiative transfer with a constant value of the gas opacity. 
This affects the density-temperature structures of the atmospheres in a decisive way: the low value chosen 
for the gas opacity in the model set used by \cite{Wacht02} results in much higher gas densities at a given 
temperature, translating into systematic differences in mass-loss rates. 
A strength of the present model grid, resulting from its more realistic input physics, is that synthetic observables 
such as photometric colours compare quite well with observations, as is discussed in the following section. 
Therefore, we conclude that using the formula by \cite{Wacht02} 
in stellar evolution models
probably leads to an overestimation of the
mass-loss. For a discussion why we have not given a new formula based on the current grid, see Sect.~4.5
of \cite{MatWH10}.

\subsection{Photometric properties}                                   
\label{s:coloureffects}

The spectra from the inner atmospheres are dominated by the usual molecular features
found in cool carbon stars -- CO, C$_2$, CN, HCN, C$_2$H$_2$, etc. If dust is formed
farther out, the spectra are modified by the amC grain opacity, attenuating the visual
spectra and adding dust emission for wavelengths $\gtrsim$ 2$\mu$m. 
This was discussed in some detail in Papers II and III.
In particular, the effects of the dust opacity for different photometric filters are discussed
and shown in Fig.~3 of Paper III.

\begin{figure}
  \resizebox{\hsize}{!}{\includegraphics{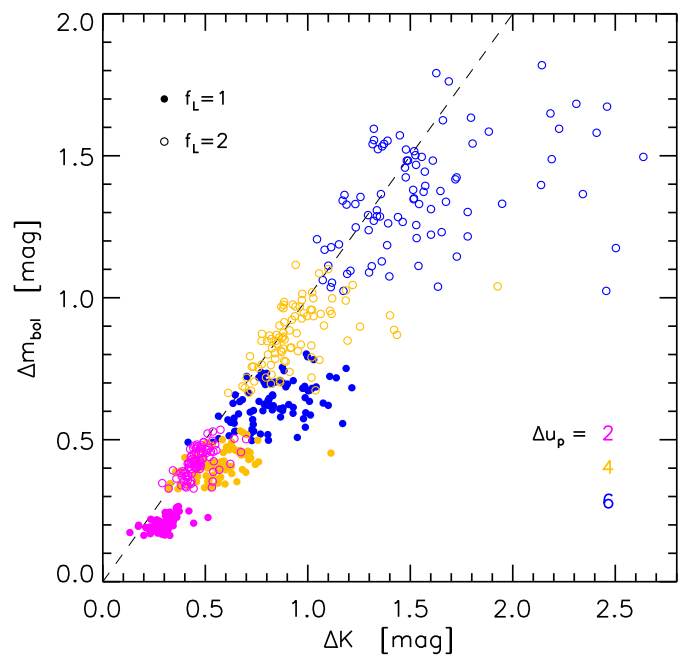}}   
  \caption{Correlation between the bolometric amplitudes, $\Delta$m$_{bol}$, and the
  amplitude in $K$, $\Delta K$, for the grid models. The dashed line is the 1:1--line.}
  \label{figmbolK}
\end{figure}

In short, the $K$ magnitude is a good proxy of the luminosity ($K$ being a filter where 
the dust extinction and emission nearly balance each other, see Fig.4 in Paper II), and thus $\Delta K$ 
is a measure of luminosity variations. This is illustrated in Fig.~\ref{figmbolK}, where these two quantities
are plotted against each other and most of the models are located close to the 1:1--line.
Models with significantly larger $\Delta K$ correspond to heavily enshrouded objects
where the entire SED is shifted towards longer wavelengths.

In contrast to the $K$ band, the $J$ filter is more affected by  dust extinction, so the \mbox{($J$\,--\,$K$)} 
colour will mostly
be a measure of the importance of dust extinction in the atmosphere/wind regions
(apart from the obvious $T_\star$ dependence).
In the following we describe how the photometric properties (i.e. amplitudes in various
filters, mean colour indices) vary across our grid of models and how they compare to observations.

\begin{figure}
  \resizebox{\hsize}{!}{\includegraphics{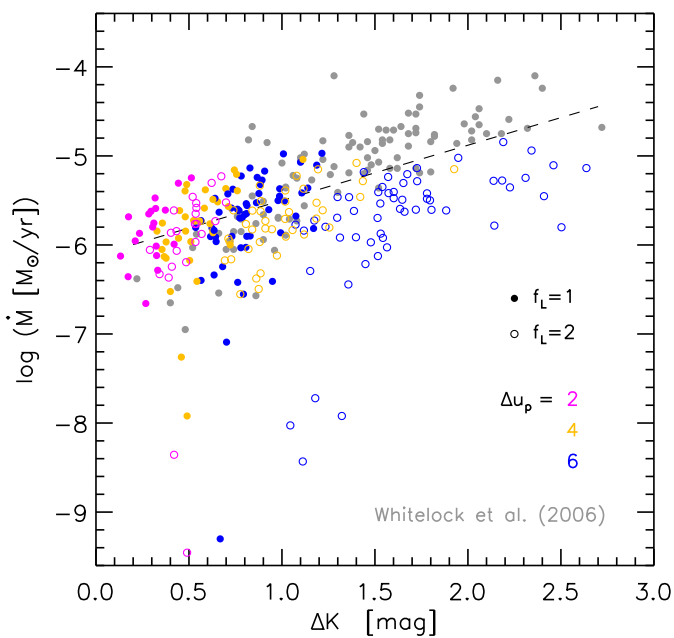}}   
  \caption{Mass-loss rates as a function of the $K$ amplitude of the models,
  as well as for observed stars in W06 (grey dots). The dashed line is a fit for carbon miras
  from \protect\cite{GroeWSK98}, Eq. 11.}
  \label{fig5}
\end{figure}

As the main observational reference we used the compilation of time-series, near-infrared (NIR) photometry 
published by Whitelock et al. (2006; below W06). 
They presented comprehensive data sets for a large sample of Galactic carbon-rich AGB stars.
Here we adopted the results for their sub-sample of well-characterized Mira variables (cf. Paper\,III).

A comparison of the mass-loss rates as a function of the $K$ amplitudes is shown 
in Fig.~\ref{fig5} for the models developing winds and for observational results taken 
from \cite{WhFMG06}, hereafter often referred to as W06.
This is our main observational reference where time-series of NIR photometry
for Galactic carbon-rich AGB stars are presented.
We note in Fig.~\ref{fig5} that the inferred very high mass-loss rates ($> 10^{-5}\  M_\odot \, yr^{-1}$)
for many (mostly distant, very red) stars have no counterparts among the models in our grid.
It is also evident from the figure that our models show a significant
spread in mass-loss rate for given $\Delta K$.
In this context, we recall that $\Delta K$ is, in essence, a measure of the luminosity 
amplitude, and that we showed earlier that this quantity {\it per se} (i.e. as far as it is not a 
consequence of a higher piston velocity amplitude but only an effect of $f_{\rm L}$) has very little 
influence on the dynamics and mass-loss rate (see Figs. \ref{clsdistr} and \ref{fig2}). 
With other stellar parameters not constrained, and no underlying pulsation model 
(giving an intrinsic coupling between the amplitudes of velocity and luminosity at 
sub-photospheric layers) 
a tight correlation between mass-loss rate and $\Delta K$ in the models is therefore not expected.

\begin{figure}
  \resizebox{\hsize}{!}{\includegraphics{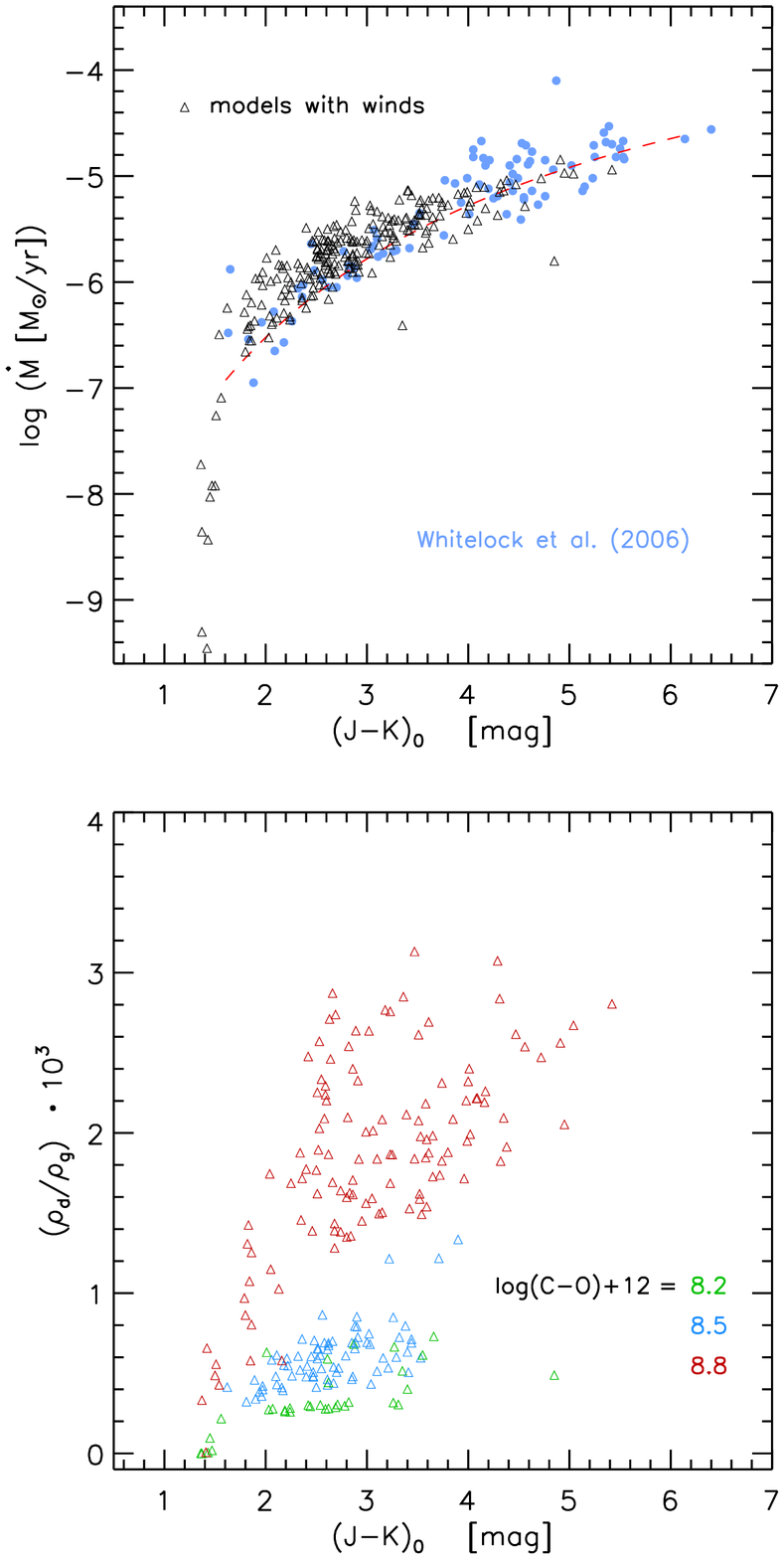} }   
  \caption{Upper panel: mass-loss rate as a function of mean \mbox{($J$\,--\,$K$)} colour 
  for models and observed stars. The dashed line is a fit from 
  \protect\cite{GGCGL12} for LMC carbon stars, the values derived by W06
  are plotted as blue dots. The lower panel shows the dust-to-gas ratio
  as a function of the mean \mbox{($J$\,--\,$K$)} colour for the models.}
  \label{fig4}
\end{figure}

How do the mass-loss rates of the  models as a function of the calculated 
mean \mbox{($J$\,--\,$K$)} colour compare with the corresponding observed/derived quantities for AGB stars? 
In Fig.~\ref{fig4}  we see that the models and observations show a similar trend and that the models cover
most of the space occupied by observed data from \cite{WhFMG06}. 
When considering the distribution of models in this diagram, one must bear in mind that the relative weights 
of parameters for the model grid do not correspond to those of  an observed sample.

\begin{figure}
  \resizebox{\hsize}{!}{\includegraphics{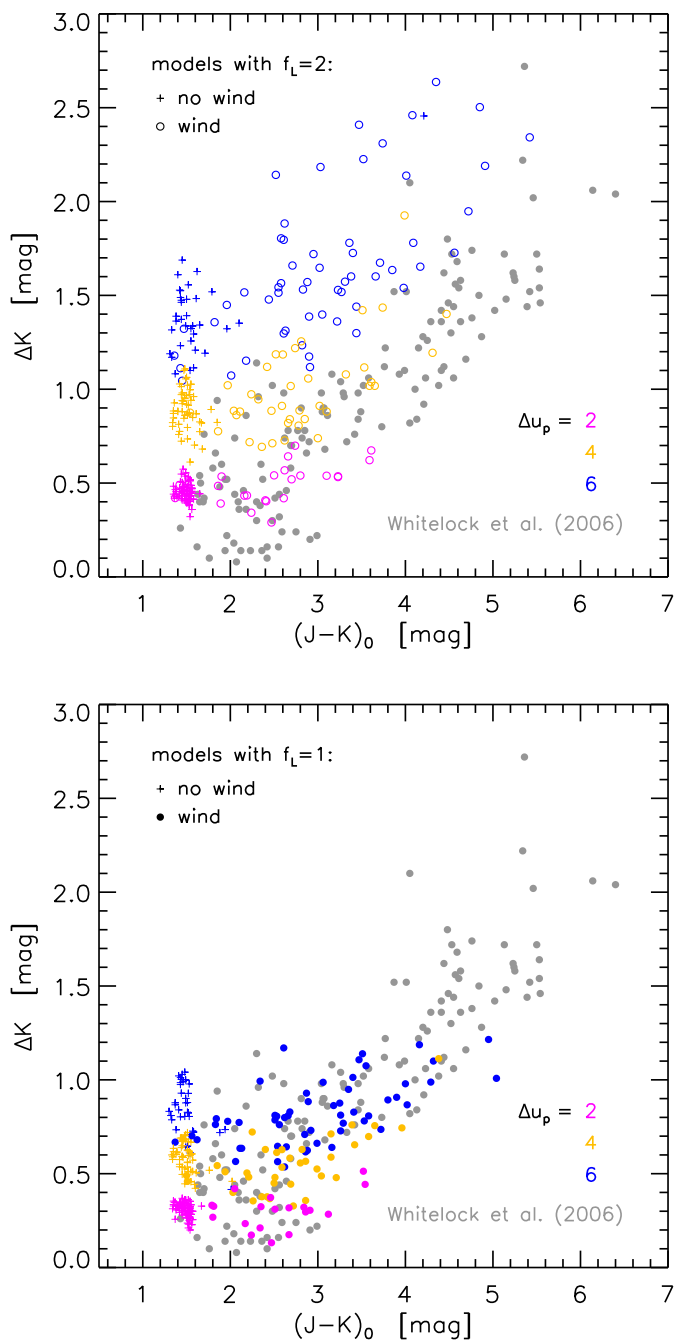}}   
  \caption{The amplitude of the $K$-magnitude variations vs. mean \mbox{($J$\,--\,$K$)} for 
   the grid models and observed values from W06 (grey dots). 
   The upper panel shows the $f_{\rm L}$=2 models, and the lower one the $f_{\rm L}$=1 ones. }
  \label{fig7}
\end{figure}

The clear correlation of mass-loss rate and \mbox{($J$\,--\,$K$)} seen in Fig.~\ref{fig4}
for both models and observations is a demonstration of the fact that this colour is a good proxy 
for the optical depth of the circumstellar envelope. 
At a closer look, however, the good quantitative agreement between models and observations 
is difficult to understand, considering the fact that the empirical values of W06 are based on the assumption 
of a constant, rather high dust-to-gas ratio that disagrees with the model results 
(see lower panel of Fig.~\ref{fig4}).
\footnote{
The observed mass-loss rates, in e.g. W06,
are derived from the measured IRAS 60\,$\mu$m flux, involving estimated distances (squared) and a 
fixed dust-to-gas ratio. This value is commonly assumed  to be around 0.005
\citep[][W06, based on \cite{Jura87}]{GGCGL12}, which is systematically higher
than in our models.
} 
The clear discrepancies in the dust-to-gas ratio must be compensated for by other assumptions 
underlying the empirical data and/or the models to arrive at the good agreement in the 
correlation between mass-loss rate and \mbox{($J$\,--\,$K$)}.

A way to avoid the uncertainties in  the mass-loss rates derived from observations 
is to compare synthetic observables of the dynamic models with directly observed quantities. 
A rather easily obtained observable is the amplitude in the $K$ magnitude,  $\Delta K$. 
The computed values of $\Delta K$ as  a function of the mean \mbox{($J$\,--\,$K$)} colour
are displayed in Fig.~\ref{fig7} together with observational data 
from W06.
Since $\Delta K$  is a measure of the luminosity variations and both $f_{\rm L}$ and $\Delta u_{\rm p}$ 
have an influence on the luminosity amplitude, we split this figure into panels for the two
different $f_{\rm L}$ values and show the $\Delta u_{\rm p}$ values in different colours.
\footnote{
Note that while both parameters affect the luminosity amplitude, the different $\Delta u_{\rm p}$ also 
correspond to different inputs of kinetic energy at the inner boundary, 
which is not the case for $f_{\rm L}$, cf. Sect~\ref{s:DMAs}.
}
The plotted value of $\Delta K$  is the mean of the amplitudes in $K$ for all selected cycles 
(i.e. at least four), whereas the \mbox{($J$\,--\,$K$)} value is the temporal mean of this colour.
We see that the models with $f_{\rm L}$ = 2 and the largest piston amplitude have photometric colours
that are rarely observed for carbon stars. 
Note also the clear dependence of $\Delta K$ on the piston amplitude, as well as the fact that 
wind models and those without a wind have approximately the same $\Delta K$ for a 
given $\Delta u_{\rm p}$ (except for models with $f_{\rm L}$=2 and a very  substantial wind, i.e. large \mbox{($J$\,--\,$K$)}).
In Paper~III  (their Fig.~11) it was illustrated that \mbox{($J$\,--\,$K$)} $\sim$ 1.6 marks the border between the 
models with and without winds; here we see that also for the present grid a \mbox{($J$\,--\,$K$)} value 
of around 1.7 marks this division.

\begin{figure}
  \resizebox{\hsize}{!}{\includegraphics{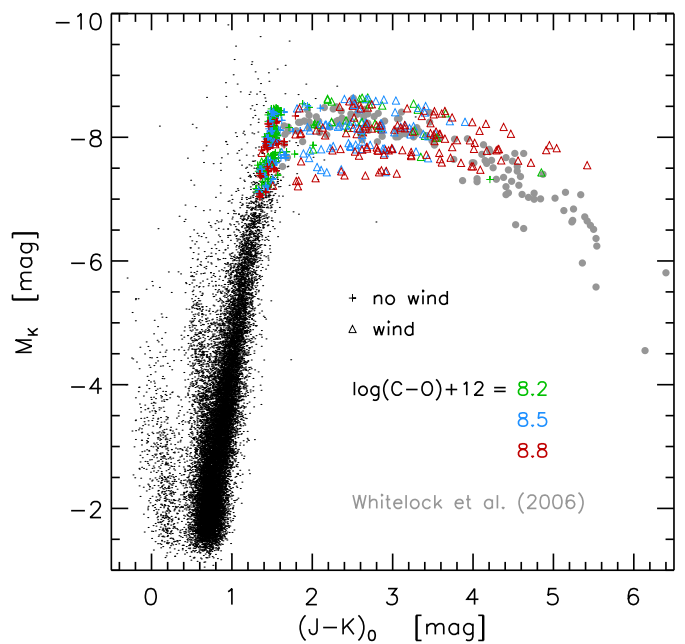}}     
  \caption{Colour-magnitude diagram showing $K$-magnitudes vs. 
  mean \mbox{($J$\,--\,$K$)} for the grid models together with
  the observed sample of W06 and 2MASS data for Baade's window (see Paper\,III for a detailed description).}
  \label{fig6}
\end{figure}

In Fig.~\ref{fig6}, showing a colour-magnitude diagram, we compare $K$ magnitude 
values from the models with observations  in Baade's window and  from W06. 
These data were  discussed in Paper~III.
Here the model values also cover most of the space occupied by the observed giant stars, 
except for the reddest objects. 
We note that models with some of our chosen parameters have no counterparts among the
observed stars, especially the low-mass, low-luminosity models that have large enough 
carbon excess and piston velocity amplitudes to initiate a wind.
As we already noted, the latter group also deviates from observations in Fig.~\ref{fig3b}.

\begin{figure}
  \resizebox{\hsize}{!}{\includegraphics{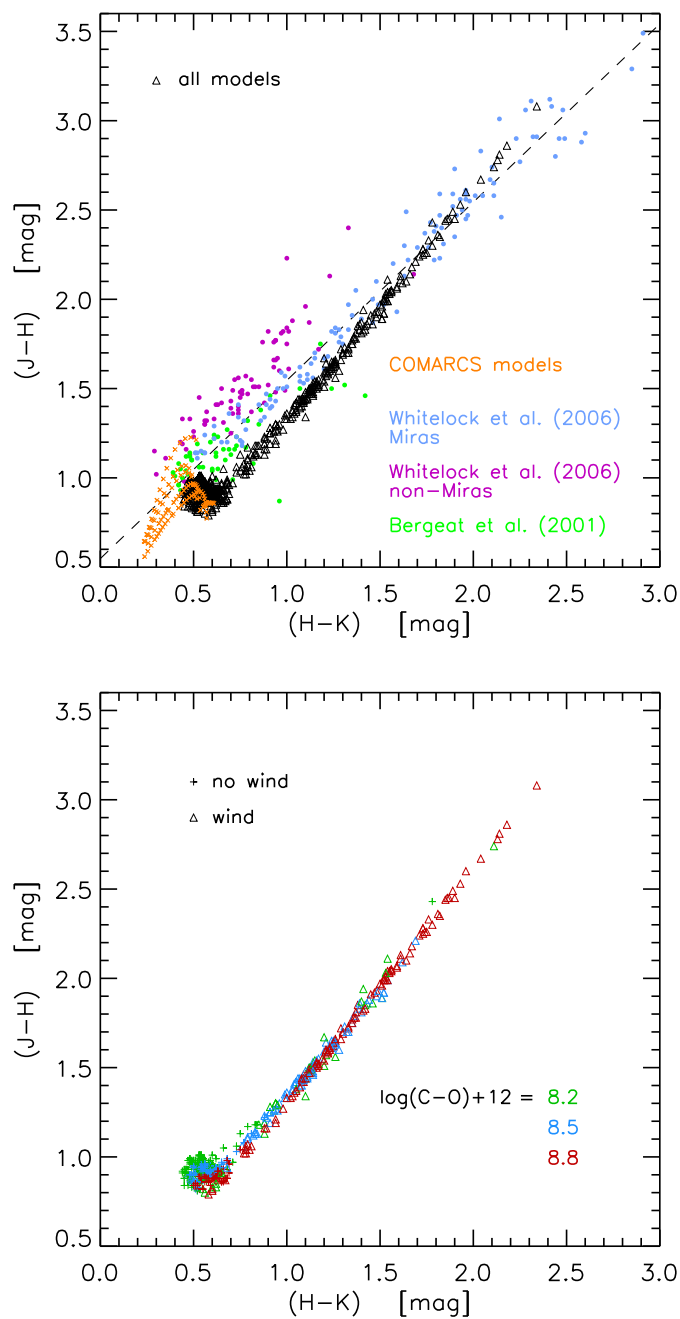}}     
  \caption{Mean \mbox{($J$\,--\,$H$)} colour vs. mean \mbox{($H$\,--\,$K$)} 
  for the model grid and some observed samples.
  The fit in the upper panel comes from W06, Eq.~2. We also plot values for hydrostatic
  COMARCS models with low (C--O), see text.} 
  \label{fig8}
\end{figure}

In the commonly used \mbox{($J$\,--\,$H$)} vs. \mbox{($H$\,--\,$K$)} plot, shown in Fig.~\ref{fig8}, we note 
a general agreement between model colours and observed ones for models with a substantial wind. 
The red \mbox{($J$\,--\,$H$)} and \mbox{($H$\,--\,$K$)} colours are due to dust 
(for details see Paper III, Sect.~5.2).
For the wind-less models we note that they appear  ``too blue'' 
in \mbox{($J$\,--\,$H$)} and/or  ``too red'' in \mbox{($H$\,--\,$K$)} compared with the observed samples. 
It is interesting to note that hydrostatic models that are computed with still smaller carbon 
excesses (corresponding to C/O ratios around 1.1, Paper~I) 
do fall around \mbox{($H$\,--\,$K$)},\mbox{($J$\,--\,$H$)} $\simeq$ (0.4, 1.1) -- 
for these stars the dynamic effects are expected to be small.

\section{Summary and conclusions}
\label{s:summary}

The results presented in this paper are based  on
state-of-the-art atmosphere and wind  models of carbon stars, spanning a 
considerable range in fundamental stellar parameters (given in Table~\ref{t:gridparam})  
with three values of carbon excess, three values of piston velocity amplitude, and two values of 
the parameter $f_{\rm L}$ defining the luminosity amplitude. 
For these 540 models, we performed {\it a posteriori} radiative transfer calculations 
for representative snapshots of structures, covering wavelengths from 0.35 to 25 $\mu$m.
We considered opacities from gas (atomic and molecular lines, continuous sources) as well as from 
amorphous carbon dust grains (when present). 
The results are available for downloading and include dynamic properties (mass-loss rates, 
outflow velocities, dust condensation degrees) as well as spectra and photometric quantities. 
The data, especially the mass-loss rates, can be used in stellar evolution modeling, or for 
population synthesis where photometric properties are needed. 
Another possibility is to study individual carbon stars using these data.

When {\em comparing our results with observed quantities} for individual or samples of carbon stars,
several aspects must be kept in mind. First we have the intrinsic limitations and assumptions in 
the models, such as neglecting 3D effects by assuming spherical symmetry, and using the small-particle 
limit for grain opacities. Second, certain parameter choices for the current modelling, such 
as the velocity and luminosity amplitudes, are proxies for a more realistic treatment of the 
pulsational behaviour of real AGB stars. Third, our grid contains
models regardless of how common a particular parameter combination would be in a real stellar 
population, characterized by an initial mass function, age, and metallicity distribution, and 
modified by stellar evolution.

Despite these caveats, we see, in general, a rather satisfactory agreement of typical models with
representative stars, regarding both dynamic and photometric properties, such as mass-loss rates vs. 
\mbox{($J$\,--\,$K$)} colours, $K$ magnitudes vs. \mbox{($J$\,--\,$K$)} colours, and 
the colour-colour diagram of \mbox{($J$\,--\,$H$)} vs. \mbox{($H$\,--\,$K$)} 
(the latter with the exception of models with small or no mass-loss; probably an effect of lacking 
low (C--O) values in the current grid). 
This is in line with our earlier detailed comparisons of selected models with individual
carbon stars \citep[see Papers II and III;][]{GaHJH04, NowHA10, SAHNPVH10}.

Two types of systematic exceptions from the good overall agreement between models and 
observations should, however, be mentioned here: in various diagrams we note 
(a) groups of models with no observed counterparts, and
(b) ranges of observed quantities not fully covered by models. 
Both phenomena can, in principle, be caused by the chosen range of parameters of the model grid, 
or by effects of intrinsic physical assumptions in the models. 
In particular, we find the following notable discrepancies:
(i) our grid does not fully cover the extreme red end of the \mbox{($J$\,--\,$K$)} range in the 
\cite{WhFMG06} sample, possibly because the highest mass-loss rates in the models 
are smaller than the highest values derived for that group of stars. 
In this context, we note that the value of the dust-to-gas ratio  
commonly assumed to derive mass-loss rates from observations (i.e. 0.005) is significantly higher
than the typical values found in our models (cf. Fig.~\ref{fig3}), and that the theoretical values
show a large spread.
(ii) Models combining low mass and low luminosity with high enough carbon excess and 
pulsation amplitude to initiate a wind have no counterparts among the observed stars in the 
diagrams showing mass loss vs. period or $K$ magnitude vs. \mbox{($J$\,--\,$K$)}. 
In addition, the models with the highest carbon excess used in this grid have a tendency to produce 
too high wind velocities compared with the bulk of observed stars. 
A major determining factor for the outflow velocity is the carbon excess, and thus the amount of 
amC grains formed. 
This leads us to conclude that stars with such high carbon excess are rare.  
(iii) In contrast to models at the high end of the assumed (C--O) range, models that do form winds 
with the lowest (C--O) values tend to show too low outflow velocities compared with their comparatively 
high mass-loss rates. 
Based on an exploratory study by \cite{MatH11}, we expect that 
using size-dependent grain opacities (instead of the small particle limit) will increase the wind speeds 
of this model group without affecting the mass-loss rates significantly, bringing them into agreement 
with the bulk of the observed stars in the mass-loss rate vs. wind velocity diagram. 

As a consequence of the present results, we plan to improve the modelling regarding both intrinsic 
model assumptions and parameter ranges. 
In particular, we will provide a full grid based on size-dependent grain opacities, which may 
both solve current problems with wind velocities, and probably improve synthetic $V$ magnitudes, for example, 
for which the effects of dust are crucial. 
Then also smaller carbon excesses will be included in the parameter range of the grid, which is expected to
bring the location of models with low or no mass loss in the 
\mbox{($J$\,--\,$H$)} vs. \mbox{($H$\,--\,$K$)} diagram into better agreement with observations.

Conclusions relevant for the {\em application of mass-loss rates in stellar evolution models} can be 
summarized as follows:
the dynamic effects resulting from increasing the luminosity amplitude relative to the grid 
by \cite{MatWH10} by a factor of 2 for otherwise identical parameters (i.e.  similar input of kinetic energy 
by pulsation)  are small, both regarding the mass-loss rates and the outflow velocities. 
Moreover, as demonstrated in the exploratory study by \cite{MatH11}, using 
size-dependent grain opacities instead of the small-particle limit probably does not affect the mass-loss 
rates considerably (despite a noticeable effect on outflow speeds for slow winds). 
Some exceptions to these conclusions may apply to marginal winds (low velocity and low 
mass-loss rates), but these are not expected to have a significant effect on evolution models. 
Therefore we conclude that the mass-loss rates given by \cite{MatWH10} 
can be used for stellar evolution modelling, 
even in the light of current and pending improvements affecting synthetic spectra, photometry, and 
other observables derived from the models.

It should also be mentioned here that the mass-loss formula given by 
\cite{Wacht02}
(based on an older generation of dynamical wind models)
gives systematically higher values of mass-loss rates for the same combinations of stellar parameters. 
In the light of the rather good agreement of our model grid with observations, mass-loss rates resulting 
from the \cite{Wacht02} formula probably need to be revised downwards.


\begin{acknowledgements}
This research was funded by the Austrian Science Fund (FWF): P21988-N16, P23006, P23586.
This work was supported by the Swedish Research Council. 
BA acknowledges funding by the contracts ASI-INAF I/016/07/0 and ASI-INAF I/009/10/0. 
This research has made use of 
(i) NASA's Astrophysics Data System, 
(ii) the VizieR catalogue access tool, CDS, Strasbourg, France,
(ii) the SIMBAD database, operated at CDS, Strasbourg, France, and
(iii) the NASA/IPAC Infrared Science Archive, which is operated by the 
Jet Propulsion Laboratory, California Institute of Technology, under 
contract with the National Aeronautics and Space Administration.
This publication makes use of data products from the Two Micron All Sky Survey, 
which is a joint project of the University of Massachusetts and the Infrared Processing 
and Analysis Center/California Institute of Technology, funded by the 
National Aeronautics and Space Administration and the National Science Foundation.
The computations were performed on resources
provided by the Swedish National Infrastructure for Computing (SNIC)
at UPPMAX. 
We thank the anonymous referee for the comments that helped improve the paper.

\end{acknowledgements}

\bibliographystyle{aa}
\citeindextrue

\bibliography{cphot4}


\begin{appendix}
\section{Details on the representative cases}
\label{a:repremods}



In Sect.~\ref{s:repcases} we defined different classes of time-dependent behaviour and illustrated them with 
the help of selected models, as shown in Fig.~\ref{fig1rev}. While we kept the stellar mass (0.75\,$M_\odot$) 
and the f$_{\rm L}$ parameter (2) fixed for this group of models, the other parameters varied
as listed in Table~\ref{t:phototypic}. 
The models are labelled as $T_\star$\,/\,log($L_\star$)\,/\,$M_\star$\,/\,log(C--O)+12\,/\,$\Delta u_{\rm p}$\,/\,$f_{\rm L}$ with L$_\star$ and M$_\star$ in solar units.

The photometric amplitudes and other properties for the illustrative cases can also be found in Table~\ref{t:phototypic}.
The column $A_V^{CSE}$ gives the extinction in the $V$ band (at 0.55\,$\mu$m)
at the phase 0.25 (or, more precisely, the mean of these values for all computed cycles); as is
further discussed in Paper~III, this is a measure of the lower limit of the dust extinction in the model.

\begin{table*}
\begin{center}
\caption{
Model parameters and resulting photometric properties of the representative cases plotted in Fig.~\ref{fig1rev}. 
The $\Delta m_{\rm bol}$,  
 $\Delta K$ and $\Delta V$ columns give the mean amplitude over the computed cycles of
 the bolometric, $K$- and $V$-band radiation (in magnitudes), whereas the ``$V$-range'' column
 gives the full range of $V$-band variations (cycle-to-cycle variations). 
The column labelled $A_V^{\rm CSE}$ gives the typical optical
 depth in the $V$ band at bolometric phase 0.25 (see text). 
 The typical positions in the two-colour \mbox{($J$\,--\,$H$)}
 vs. \mbox{($H$\,--\,$K$)} diagram are also given. }
\begin{tabular}{clcccccr@{,}l}     
\hline
\hline
 class & model & $\Delta m_{\rm bol}$ & $\Delta K$  & $\Delta V$ & $A_V^{\rm CSE}$  & $V$-range & \mbox{($H$\,--\,$K$)}\ \ &\ \mbox{($J$\,--\,$H$)} \bigstrut[t] \\
& $T_\star$\,/\,log($L_\star$)\,/\,$M_\star$\,/\,log(C--O)+12\,/\,$\Delta u_{\rm p}$\,/\,f$_{\rm L}$ & [mag] & [mag] & [mag] & [mag] & [mag] & \mbox{[mag]}\ \ &\ \mbox{[mag]} \bigstrut[t] \bigstrut[b]\\       
       
       \bigstrut[b]\\
\hline
 {\it pp} & 3000\,/\,3.55\,/\,0.75\,/\,8.50\,/\,2\,/\,2.0 & 0.38  &  0.42 &  0.72  &  0.00  & 0.72?  &  0.53\  &\   0.92  \bigstrut[t] \\
 {\it pm} & 2800\,/\,3.55\,/\,0.75\,/\,8.50\,/\,2\,/\,2.0 & 0.34  &  0.40 &  0.84  &  0.05  & 0.84?  &  0.58\   &\   0.91  \\
 {\it pn} & 2600\,/\,3.70\,/\,0.75\,/\,8.20\,/\,4\,/\,2.0 & 0.69  &  0.68 &  1.33  &   0.56  & 2.34  &  0.6--0.9\   &\   0.7--1.2 \bigstrut[b]\\
 \hline
 {\it we} & 2800\,/\,3.85\,/\,0.75\,/\,8.20\,/\,6\,/\,2.0 & 1.42  &  1.48 &  2.37  &  2.71   & 4.67  &  0.9--1.4\   &\   0.9--1.8  \bigstrut[t]\\
 {\it ws} & 2800\,/\,3.85\,/\,0.75\,/\,8.50\,/\,2\,/\,2.0 & 0.45  &  0.43 &  0.69  &  1.82   & 1.00  &  0.91\   &\   1.25   \\
 {\it wp} & 2800\,/\,3.70\,/\,0.75\,/\,8.80\,/\,4\,/\,2.0 & 0.81  &  0.84 &  3.78  &  2.98    & 4.46  &  1.1--1.4\   &\   1.3--1.8 \\
 {\it wn} & 2800\,/\,3.70\,/\,0.75\,/\,8.80\,/\,6\,/\,2.0 & 1.26  &  1.53 &  6.71  &  4.55   & 8.94  & 1.2--1.8\   &\   1.8--2.4  \bigstrut[b]  \\
\hline
\end{tabular}
\label{t:phototypic}
\end{center}
\end{table*}

\subsubsection*{{\it pp}:  3000\,/\,3.55\,/\,0.75\,/\,8.50\,/\,2\,/\,2.0} 
In this particular case, the largest extension is 1.5 stellar radii.
No dust is formed in this model, the light curves show a very regular and repeated variation from
cycle to cycle; the amplitudes (maximum-to-minimum) are 0.38$^{\rm mag}$, 0.42$^{\rm mag}$ 
and 0.72$^{\rm mag}$ for the
bolometric, $K$ and $V$ magnitudes. It has the position (0.53, 0.92) in the \mbox{($J$\,--\,$H$)} vs. \mbox{($H$\,--\,$K$)} two-colour-diagram.

\subsubsection*{{\it pm}: 2800\,/\,3.55\,/\,0.75\,/\,8.50\,/\,2\,/\,2.0}
This model has the same parameters as the pp one except that it is 200\,K cooler. 
It is classified as {\it pm} and
shows a different dynamic behaviour where the outer boundary varies from smallest to
largest extension and back in almost precisely two luminosity periods. 
The largest extension of the atmosphere is about 2 $R_\star$; it extends to $\gtrsim 2 R_\star$ every fourth (luminosity) cycle interleaved with an expansion to 
$\lesssim 2 R_\star$ (also every fourth cycle). This behaviour is coupled to a higher dust
condensation every fourth cycle when the stellar material reaches greater heights. We can note that
although the condensation degree is lower than 10$^{-3}$, the influence of the dust on the
photometric behaviour is clearly visible. The amplitudes are 0.34$^{\rm mag}$, 0.40$^{\rm mag}$ 
and 0.84$^{\rm mag}$ for the 
bolometric, $K$ and $V$ magnitudes, respectively, where we note that the amplitude in $V$ has
increased (from the previous {\it pp} model) because of the dust. 
The minima (and maxima) are fainter in the presence of dust a distance, r, 
$\sim$ 1.5 -- 2 $R_\star$. 
[It might be classified as a SRa type variable star.]
Its position in the \mbox{($J$\,--\,$H$)} vs. \mbox{($H$\,--\,$K$)} diagram is at (0.58, 0.91), that is,
slightly redder in \mbox{($H$\,--\,$K$)} than  the previous one.

\subsubsection*{{\it pn}: 2600\,/\,3.70\,/\,0.75\,/\,8.20\,/\,4\,/\,2.0}
Compared with the {\it pm} model, it has a lower $T_\star$ by 200\,K and an increased luminosity,
and also an increased piston velocity amplitude but a lower carbon excess.
Here the pulsations cause the atmosphere to expand to between
2 and 4 $R_\star$ in a rather irregular way. The condensation degree varies from
a few percent up to about 25\% without any obvious correlation to the pulsation behaviour of the atmosphere.
Now we see a stronger influence of dust on the photometric properties. The amplitudes of the
variation in bolometric, $K$ and $V$ magnitudes are 0.69$^{\rm mag}$, 0.68$^{\rm mag}$ 
and 1.33$^{\rm mag}$ (if we take the mean over all computed cycles; the maximum-to-minimum 
variation in $V$ over the computed range is 2.34$^{\rm mag}$).
[It might be classified as a mira variable.]
This model is situated around (0.6--0.9, 0.7--1.2) in a \mbox{($J$\,--\,$H$)} vs \mbox{($H$\,--\,$K$)} 
plot, that is, redder than the {\it pm} model, especially in \mbox{($H$\,--\,$K$)}.

\subsubsection*{{\it we}: 2800\,/\,3.85\,/\,0.75\,/\,8.20\,/\,6\,/\,2.0}
This model is 200\,K warmer than the {\it pn} one, but with increased luminosity
and piston velocity amplitude.
This model rarely has an atmospheric extension smaller than 5 $R_\star$. 
The condensation degree varies between 5 and 30\%, which implies quite a strong effect
on the photometric properties. The amplitudes for the bolometric, $K$ and $V$ magnitudes
are 1.42$^{\rm mag}$, 1.48$^{\rm mag}$ and 2.37$^{\rm mag}$ (the total range in $V$ over all the 
computed cycles is 4.67$^{\rm mag}$).
[It might be classified as a mira variable during its wind phase.]
In the two-colour \mbox{($J$\,--\,$H$)} vs \mbox{($H$\,--\,$K$)} diagram the model occupies 
the range 0.9--1.4 in \mbox{($H$\,--\,$K$)} and 0.9--1.8 in \mbox{($J$\,--\,$H$)}; compared with
the {\it pn} model, it has moved along the blackbody line towards cooler temperatures.

\subsubsection*{{\it ws}: 2800\,/\,3.85\,/\,0.75\,/\,8.50\,/\,2\,/\,2.0}
This model
has the same temperature and luminosity as the {\it we} model, but a higher carbon excess, 
so it develops a wind even at  the low piston amplitude of $\Delta u_{\rm p}$ = 2 km s$^{-1}$. 
It has a dust condensation degree that is almost constant in time, at 15$\pm$1\%. 
The amplitudes in the bolometric, $K$ and $V$ magnitudes
are 0.45$^{\rm mag}$, 0.43$^{\rm mag}$ and 0.69$^{\rm mag}$, respectively, 
(the total range in $V$ over all the  computed cycles is 1.00$^{\rm mag}$). 
Its position in the \mbox{($J$\,--\,$H$)} vs. \mbox{($H$\,--\,$K$)} diagram is at (0.91, 1.25) with a small
variation since the amount of dust is almost constant.

\subsubsection*{{\it wp}: 2800\,/\,3.70\,/\,0.75\,/\,8.80\,/\,4\,/\,2.0}
A model with
a lower luminosity than our {\it we} case but a higher carbon excess and piston 
velocity amplitude, it has a wind in which dust condenses at r  $\sim$ 1.5 $R_\star$ every 
cycle around phase 0.5.
The photometric amplitudes are 0.81$^{\rm mag}$, 0.84$^{\rm mag}$ and 3.78$^{\rm mag}$ for the 
bolometric, $K$ and $V$ magnitudes (the total range in $V$ over all the computed cycles is 
4.46$^{\rm mag}$). 
[It might certainly be classified as a mira variable.] 
The dust condensing around phase 0.5 makes the minima in $V$ quite deep. 
In the two colour diagram the model varies (loops), along the blackbody line, between 1.1--1.4 
in \mbox{($H$\,--\,$K$)} and between 1.3--1.8 in \mbox{($J$\,--\,$H$)}.

\subsubsection*{{\it wn}: 2800\,/\,3.70\,/\,0.75\,/\,8.80\,/\,6\,/\,2.0}
It has the same parameters as the {\it wp} case, but a higher piston velocity amplitude. 
Here the wind is more irregular, the dust is still formed around r $\sim$ 1.5 $R_\star$ but in a 
more irregular fashion.
The amplitudes in the bolometric, $K$ and $V$ magnitudes are 1.26$^{\rm mag}$, 1.53$^{\rm mag}$ and 
6.71$^{\rm mag}$, respectively, (the total range in $V$ over all the computed cycles is 
8.94$^{\rm mag}$). 
In the two-colour diagram the model moves during the pulsations along the blackbody line between
1.2--1.8 in \mbox{($H$\,--\,$K$)} and between 1.8--2.4 in \mbox{($J$\,--\,$H$)}.

\Online

\section{Model overview}
\label{a:overviewdata}

In the following pages we present a table with photometric and dynamic properties of
the models in the present grid.
The models are arranged in increasing effective temperature, luminosity, and stellar mass.
For each such combination the data are ordered by increasing carbon excess, piston velocity
amplitude, and $f_{\rm L}$.
In each line,
after the model parameters we list the assigned class, log~g (surface gravity in cgs units).
Then come dynamic quantities evaluated at the outer boundary:  mass-loss rate
(in solar masses per year),
the wind velocity (km/s), the carbon condensation degree, and the dust-to-gas ratio.
Note that all given values are temporal means, see Sect.~\ref{s:DMAdyn}.
Then follow the photometric properties: the (full)
amplitude of the bolometric magnitude, the mean $V$ magnitude, the range of $V$ magnitudes, the mean
$K$ magnitude and its range, and finally the colours \mbox{($V$\,--\,$I$)}, \mbox{($V$\,--\,$K$)},
\mbox{($J$\,--\,$H$)} and \mbox{($H$\,--\,$K$)}.

The luminosities and stellar masses are given in solar units. 
The carbon excesses, log(C--O)+12
 are given on the scale where 
log N$_H$ $\equiv$12.00. 
The piston velocity amplitudes, $\Delta u_{\rm p}$, are given in km/s. The $f_{\rm L}$ and class designations are
described in Sect.~\ref{s:DMAs} and Sect.~\ref{s:repcases} respectively.
All the  photometric quantities are given in magnitudes.

\section{Material at the CDS}
\label{a:download}

The resulting spectra and synthetic photometric magnitudes in various filters are available
for downloading,  some photometric data are also summarized in 
Table~B.1  
along with dynamic data, such as mass-loss rates and wind velocities.

\begin{figure*}
\sidecaption
  \includegraphics[width=12cm]{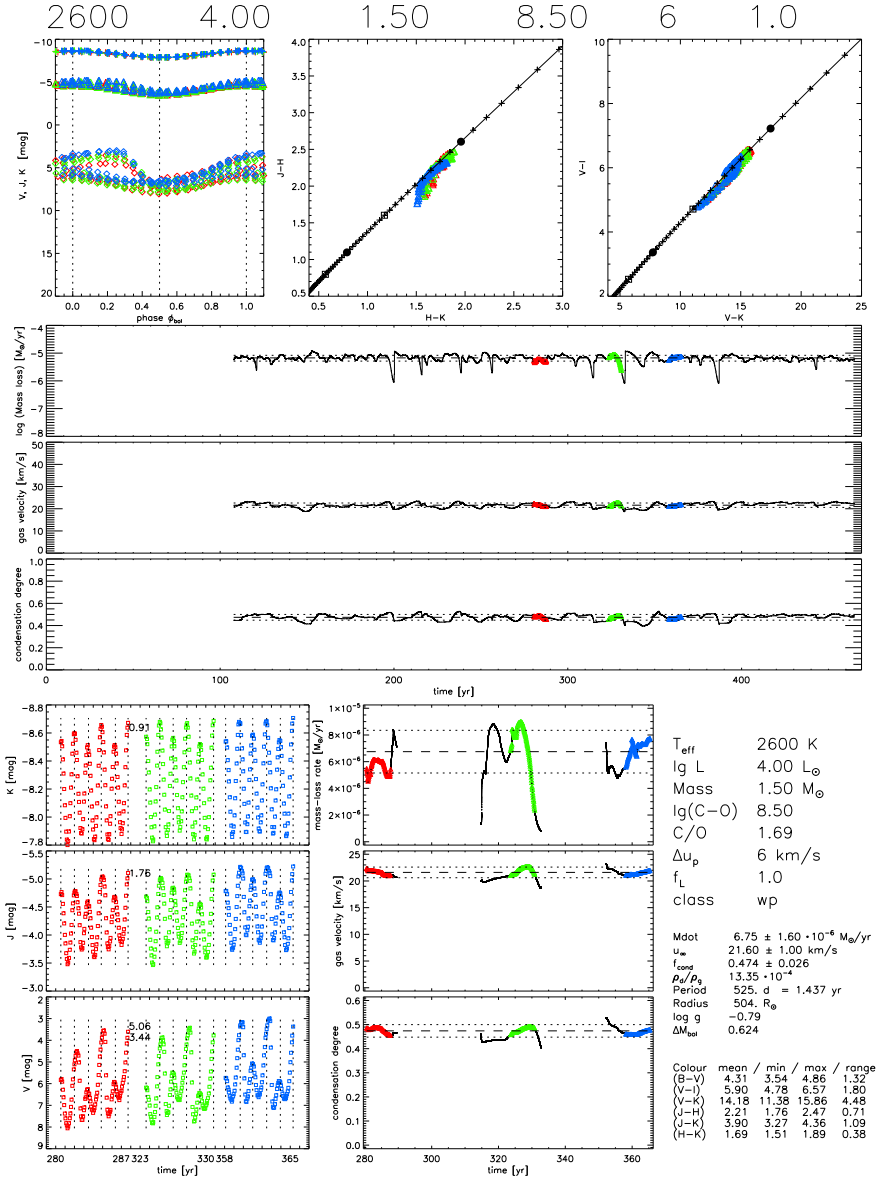}   
     \caption{Example of the first page of the fact sheets. The top label gives 
     $T_\star$, log\,$L_\star$, $M_\star$, log(C--O)+12, $\Delta u_{\rm p}$, and $f_{\rm L}$ for a quick overview. 
     The upper leftmost diagram gives the $K$, $J$ and $V$ magnitudes
     as a function of phase (with maximum luminosity at phase 0.0). The ranges are identical
     on all fact sheets to facilitate comparisons of different models; this is true for 
     the three upper diagrams. The different colours correspond to the different epochs selected for 
     the spectral synthesis. The middle and right upper panels show 
     two colour-colour diagrams, the blackbody line is also shown. 
     The three middle panels show, as a function of time, the mass-loss rate, the
     gas velocity, and the condensation degree for carbon, all three at the outer boundary (25\,$R_\star$);
     here the vertical scales are identical on all fact sheets, the time intervals do vary from model to model.
     For the models without a wind, the position of the outer boundary is shown instead of the mass-loss
     rate. For the episodic models, we show both of them. The horizontal dashed and dotted lines denote
     the mean value, and the mean value $\pm$ the standard deviation. The bottom part contains the
     light-curves in $K$, $J$ and $V$; here the dotted vertical lines denote the positions 
     of the luminosity maxima,
     the maximum range in magnitudes is also given, for the $V$ magnitude also the mean amplitude over
     a (luminosity) cycle. Enlarged versions of the three middle plots, now with automatic scaling, 
     are shown for the selected epochs. Finally, a summary of parameters, dynamic and photometric
     properties are listed.}
     \label{factsheetfig}
\end{figure*}

\begin{figure*}
\sidecaption
  \includegraphics[width=12cm]{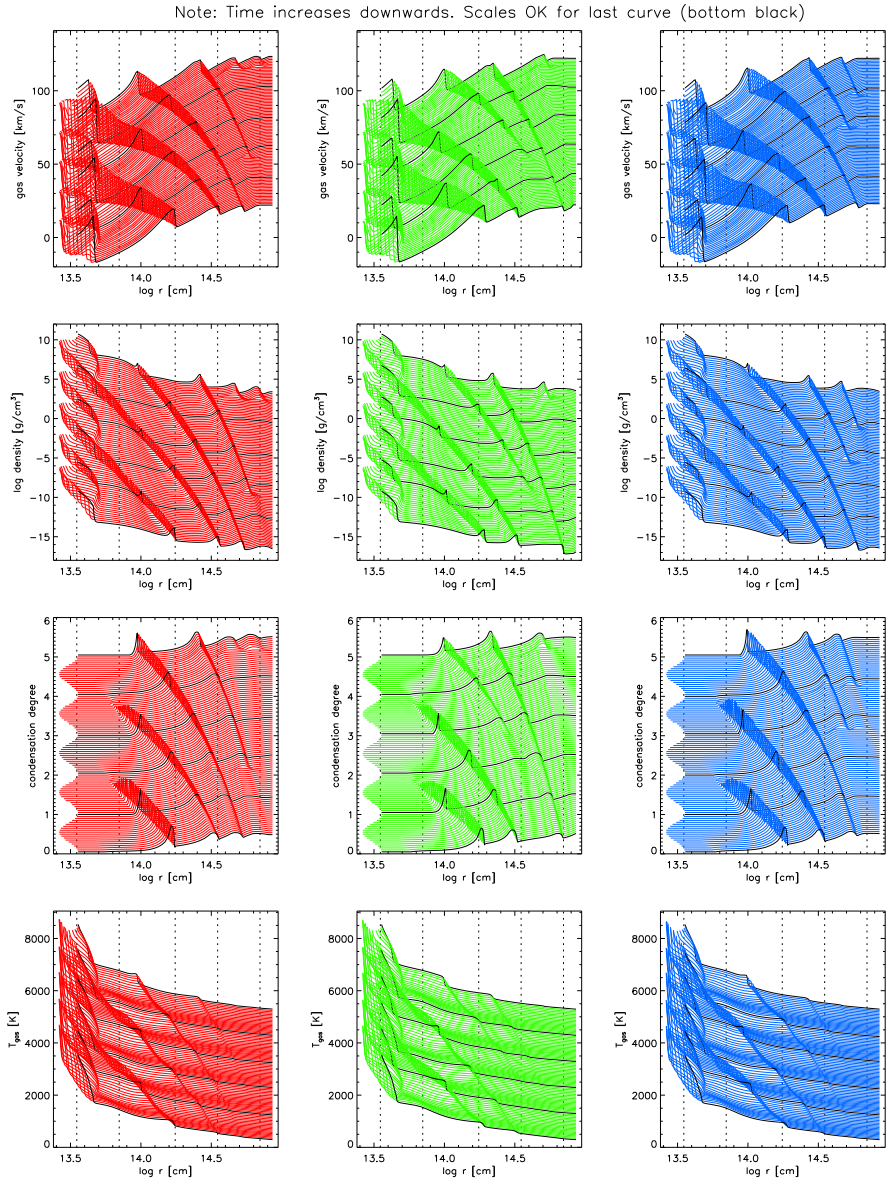}   
     \caption{Example of the second page of the fact sheets.The colours in the plots correspond to
     the selected epochs as given on the first page. In each column we show, from top to bottom, the
     gas velocity, density, carbon condensation degree, and gas temperature, all as a function of distance from
     the stellar centre. In each panel, individual curves correspond to snapshots in time, and time increases
     downwards. The distance between two curves is approximately 0.05 in phase. The curves closest
     in time to phase 0.0 are drawn with thick black lines. The ordinate scales apply to the bottom (last) curve. 
     The vertical short-dashed lines in the panels denote the distances 1, 2, 5, 10, and 20 stellar radii. }
     \label{factsheetfig2}
\end{figure*}

Among the downloadable material there is also,
for each individual model,  a two-page fact sheet that looks like Figs.~\ref{factsheetfig}
and \ref{factsheetfig2} summarizing the dynamic and photometric behaviour.
It should be noted that the values for mass-loss rates, gas velocities, and carbon condensation
degrees plotted on the fact sheets are values for each snapshot, whereas the values for these 
quantities given in the figures and tables in this article are computed as averages over
pulsation periods as described in Sect.~\ref{s:DMAdyn}.

\end{appendix}

\end{document}